\def\numcol{1}
\if\numcol1
    
    \documentclass[12pt,draftclsnofoot,onecolumn]{IEEEtran}
\else
    \documentclass[lettersize,journal]{IEEEtran}
\fi
\usepackage{amsmath,amsfonts}
\usepackage{algorithm}
\usepackage{algorithmicx}
\usepackage{algpseudocode}
\usepackage{array}
\usepackage[caption=false,font=normalsize,labelfont=sf,textfont=sf]{subfig}
\usepackage{textcomp}
\usepackage{stfloats}
\usepackage{url}
\usepackage{verbatim}
\usepackage{graphicx}
\usepackage{booktabs}
\usepackage{threeparttable}
\usepackage{adjustbox}
\usepackage{multirow}
\usepackage{cite}
\usepackage{xcolor}
\definecolor{TodoColor}{HTML}{5DBA4A}
\definecolor{TitleBlue}{HTML}{0048b1}

\begin{document}

\title{Temporal Channel Estimation for \\Generalized CSI Feedback}

\author{
    Minwoo Kim,
    Hyeonsu Lyu,
    Sehyun Ryu,
    Sojeong Park
    and Hyun Jong Yang,~\IEEEmembership{Senior Member,~IEEE}
\thanks{This paper was produced by the IEEE Publication Technology Group. They are in Piscataway, NJ.}
\thanks{Manuscript received April 19, 2021; revised August 16, 2021.}
\thanks{\textit{(Corresponding author: Hyun Jong Yang.)}}
\thanks{
    M. Kim, S. Ryu and S. Park are with the Department of Electrical Engineering, Pohang University of Science and Technology (POSTECH), Pohang 37673, South Korea
    (email: mwkim0210@postech.ac.kr; sh.ryu@postech.ac.kr).
    
    H. Lyu and H. J. Yang are with the Institute of New Media and Communications, Seoul 08826, South Korea (e-mail: hs.lyu@snu.ac.kr, hjyang@snu.ac.kr
    
    H. J. Yang is with the Dept. of Electrical and Computer Engineering, Seoul National University, Seoul 08826, South Korea (email: hjyang@snu.ac.kr).
    }
}

\if\numcol1
\else
    \markboth{Journal of \LaTeX\ Class Files,~Vol.~14, No.~8, August~2021}%
    {Shell \MakeLowercase{\textit{et al.}}: A Sample Article Using IEEEtran.cls for IEEE Journals}
\fi

\maketitle

\begin{abstract}
Efficient Channel State Information (CSI) feedback is indispensable for frequency division duplex (FDD) massive multiple-input multiple-output (MIMO) systems.
Existing compressed sensing (CS) algorithms exploit delay-domain sparsity but suffer from prohibitive iterative latency and discrete grid mismatch.
Conversely, deep learning (DL) approaches achieve rapid inference but lack spatial scalability and domain adaptability, failing to generalize to unseen propagation environments, and demand computationally heavy encoders and decoder.
In this paper, we propose TAP, a Tap-Assisted Parametric CSI Compression.
TAP is a one-shot neural framework that unifies the speed of DL with the mathematical interpretability of CS.
TAP replaces iterative pursuit with a lightweight 1D neural network that extracts dominant continuous propagation delays from temporal channel sequences via a differentiable sub-grid interpolation operator.
TAP achieves true architecture independence, enabling zero-shot generalization across diverse array geometries and unseen propagation environments.
Furthermore, TAP yields a completely decoder-free payload, allowing the BS to reconstruct the channel via a simple inverse fast Fourier transform (IFFT).
Extensive evaluations across five 3GPP environments demonstrate that TAP achieves a 3.13 to 12.22~dB channel frequency response normalized mean square error (CFR-NMSE) improvement over CsiNet while shrinking the model footprint by 660 times to under 1~MB. Operating with sub-millisecond latencies, TAP accelerates inference by 2700 times over classical iterative OMP, providing a scalable and deployment-ready solution for next-generation networks.
\end{abstract}

\begin{IEEEkeywords}
Channel State Information (CSI) feedback, compressed sensing, orthogonal matching pursuit, deep learning, zero-shot domain adaptation, channel estimation
\end{IEEEkeywords}

\section{Introduction}

Massive multiple-input multiple-output (MIMO) technologies are fundamental to achieving the high spectral efficiencies demanded by 5G and 6G networks.
In frequency division duplex (FDD) systems, the base station (BS) requires accurate downlink channel state information (CSI) to perform spatial multiplexing and precoding~\cite{Shen16-TVT}.
The user equipment (UE) must estimate the downlink channel frequency response (CFR), compress it, and feed it back over a rate-limited uplink control channel.
As BS antenna counts scale to hundreds of elements, the dimensionality of the full CFR grows proportionally with the product of transmit antennas and subcarriers.
Efficient CSI compression is therefore indispensable for the practical deployment of FDD massive MIMO.

Two broad paradigms have been studied to address this feedback bottleneck.
Classical compressed sensing (CS) algorithms such as orthogonal matching pursuit (OMP), LASSO, and TVAL3 exploit the physical delay-domain sparsity of wireless channels to recover the CSI from a small number of measurements~\cite{Pati93-ACSSC, Tropp07-TIT, Qi15-ICC, Huang17-Access, Liang20-TVT, Li13-TVAL3}.
However, their performance depends critically on the sparsity assumption, which is often only approximately satisfied in practical propagation environments, motivating the development of data-driven deep learning (DL) approaches~\cite{Yang19-MLSP}.
DL-based approaches treat CSI matrices as images and learn end-to-end encoder-to-decoder mappings, with architectures such as CsiNet~\cite{Wen18-WCL} achieving high compression ratios (CRs) by operating on truncated delay-angular domain representations.
In communication standards, 3GPP Type-1 and Type-2 codebooks offer a third, lightweight alternative but typically plateau at high normalized mean square error (NMSE)~\cite{3gpp.38.214}.

However, these existing approaches suffer from a few fundamental limitations that prevent their practical deployment.
First, CS-based methods are computationally prohibitive for real-time applications.
Recovering time-domain channel using standard OMP requires iterative large-scale inverse fast Fourier transforms (IFFTs), spatial correlations, and least squares (LS) projections, resulting in large runtimes, exceeding the sub-millisecond latency budget of 5G New Radio (NR).
Second, although many DL-based methods achieve the latency constraint, they lack both spatial scalability and cross-domain generalization as their architectures are hard-wired to fixed antenna dimensions
Therefore, any change in BS array geometry requires redesigning and retraining the network.
Moreover, because their learned latent spaces implicitly encode the propagation statistics of the training environment, models trained in one scenario (e.g., an indoor factory) suffer severe degradation in another (e.g., an outdoor macrocell).
The dense convolutional or fully connected layers in these models also scale with the antenna count, producing model footprints of hundreds of megabytes (MBs) that are impractical for memory- and power-constrained UEs.

No existing methodology simultaneously satisfies the four requirements of a practical CSI compressor: (i) sub-millisecond inference latency, (ii) a sub-megabyte model footprint, (iii) high compression fidelity across diverse propagation environments, and (iv) zero-shot generalization to unseen deployments and arbitrary antenna geometries.
The root cause of this gap is that DL methods discard the underlying physical channel structure, while CS methods fail to leverage temporal priors to accelerate inference.
A hybrid framework that embeds physical sparse-recovery structure into a lightweight neural pipeline is therefore needed to satisfy all four constraints.

\begin{figure*}[!ht]
    \centering
    \includegraphics[width=\if\numcol1 0.7 \else 0.7\fi\linewidth]{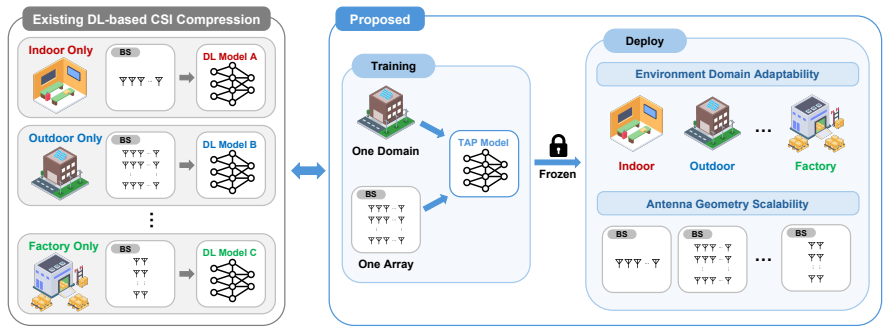}
    \caption{Comparison of previous DL-based CSI compression methods and the proposed TAP pipeline.}
    \label{fig:system_model}
\end{figure*}

To bridge this gap, this paper proposes \emph{TAP}: a Tap-Assisted Parametric CSI Compression, a lightweight, one-shot neural framework featuring antenna scalability and domain adaptability.
To the best of the authors' knowledge, TAP is the first framework to simultaneously satisfy the four requirements of a practical CSI compressor.
By unifying physical sparsity with neural efficiency, TAP achieves favorable trade-offs across compression quality, speed, compactness, and generalizability.
We have illustrated the difference between previous approaches and TAP in Fig.~\ref{fig:system_model}.
The main contributions are as follows:
\begin{itemize}
    \item \textbf{One-Shot Neural Delay Estimation:} We introduce DelayNet that predicts dominant propagation delays simultaneously from a temporal window of historical CFR sequences. This replaces the computationally expensive iterative atom selection of classical OMP with a single forward pass.
    \item \textbf{Differentiable Sub-Grid Delay Extraction:} We propose a symmetric linear power-weighted center-of-mass extractor to estimate continuous, fractional delays from discrete network outputs. This differentiable operator eliminates the dictionary mismatch problem inherent to discrete-grid OMP and enables the entire pipeline to be trained end-to-end.
    \item \textbf{Antenna-Independent Architecture:} By employing depthwise separable convolutions with antenna-averaging in the delay predictor, the proposed framework keeps the number of model parameters constant regardless of the BS antenna array size. This structural design enables zero-shot generalization across diverse antenna geometries.
    \item \textbf{Zero-Shot Cross-Domain Adaptation:} We demonstrate that TAP can generalize across entirely different propagation environments without any retraining, effectively decoupling the model from environment-specific statistics.
    \item \textbf{Decoder-Free BS Reconstruction:} Unlike existing deep learning-based feedback schemes that require executing complex neural decoders at the base station, TAP physically structures the feedback payload such that the BS can reconstruct the full CSI matrix using only a simple matrix multiplication equivalent to an inverse Fourier transform. This eliminates BS-side neural inference overhead and deployment complexity.
    \item \textbf{Comprehensive Evaluation:} Extensive evaluations across five 3GPP environments---including both ray-traced (Sionna RT and DeepMIMO) and stochastic (Clustered Delay Line, CDL) datasets---demonstrate that TAP achieves a 12 to 22~dB CFR NMSE improvement over CsiNet architecture while utilizing a 660 times smaller model size of under $1$~MB. TAP accelerates inference by 2700 times compared to iterative OMP.
\end{itemize}

\section{Related Works}

Prior work on CSI compression generally falls into DL-based and CS-based paradigms.
While each approach shows good performance along specific axes, existing methods fail to unify inference speed, reconstruction accuracy, model compactness, and zero-shot generalization.

\subsection{Deep Learning-Based CSI Feedback}
The DL-based CSI feedback literature has progressively improved reconstruction quality through architectural innovations.
The CsiNet architecture demonstrated that a lightweight convolutional autoencoder could compress truncated delay-angle CSI matrices and achieve competitive reconstruction performance~\cite{Wen18-WCL}.
Successive frameworks have enhanced fidelity by incorporating multi-resolution architectures~\cite{Lu20-ICC}, long short-term memory (LSTM) networks for temporal fusion~\cite{Wang19-WCL}, and attention mechanisms ~\cite{Xu21-WOCC, Cui22-WCL, Liao25-TITS}.
Parametric channel modeling methods have mapped heavy networks directly to physical propagation parameters~\cite{Ju24-TWC}, while model-driven deep unfolding architectures unrolled iterative solvers to neural layers~\cite{Kang22-JSAC}.
Data-driven line-spectrum frameworks have explored extracting continuous frequencies from multi-sinusoidal mixtures~\cite{Izacard19-NeurIPS}, though these lack the multi-dimensional scalability required for massive MIMO.

Despite the progress in reconstruction quality, DL-based approaches suffer from three limitations that prevent universal deployment.
First, their architectures are hard-wired to fixed spatial dimensions: the fully connected or convolutional layers are coupled to specific antenna layouts and subcarrier groupings, so any change in the BS antenna geometry requires a full redesign and retraining.
Second, the learned latent spaces implicitly encode the propagation statistics of the training environment, leading to severe performance degradation when deployed in unseen scenarios.
Third, the dense layers scale with the antenna count, resulting in model footprints that can reach hundreds of megabytes, which are impractical for memory- and power-constrained UE.

\subsection{CS-Based CSI Feedback}
Classical CS algorithms, including OMP, LASSO, and TVAL3, have been extensively investigated for CSI compression~\cite{Pati93-ACSSC, Tropp07-TIT, Qi15-ICC, Huang17-Access, Liang20-TVT}.
These approaches leverage the physical property that wireless channels are inherently sparse in the delay domain, using measurement matrices and recovery algorithms to reconstruct the channel from few observations~\cite{Choi17-CST, Shen16-TVT, Kuo12-WCNC}.
Advanced variants further exploit spatial correlation across multiple antennas via joint-sparsity OMP and Bayesian CS approaches~\cite{Baraniuk10-TIT, Ji08-TSP}, while other frameworks integrate both spatially common sparsity and temporal channel correlation to enable closed-loop sparsity-adaptive tracking~\cite{Gao15-TSP}.

However, CS-based methods face two structural limitations.
First, the iterative algorithms (e.g., OMP, LASSO) scale poorly with dictionary size and the number of resolvable paths, producing latencies on the order of seconds per feedback instance, which is three to four orders of magnitude beyond the sub-millisecond budget of 5G NR~\cite{He25-TVT}.
Second, standard discrete dictionaries assume that physical propagation delays align with predefined grid points.
When true delays fall between these grid points, the resulting basis mismatch degrades both support detection and amplitude recovery accuracy.
These computational and structural constraints prevent CS methods from meeting real-time deployment requirements.

\subsection{Domain Adaptation and Generalization}
The practical deployment challenge of environment mismatch has motivated extensive domain-adaptation research in CSI feedback~\cite{Zeng21-TCCN}.
When a DL model trained in one environment is evaluated in another, the mismatch in propagation statistics degrades performance~\cite{Feng23-ICC}.
To mitigate this, researchers have proposed transfer learning via fine-tuning~\cite{Zeng21-TCCN, Sattari25-SPAWC}, model-agnostic meta-learning~\cite{Xiao23-ICC}, domain adaptation with limited target data~\cite{Liu24-TWC}, and domain-adversarial training~\cite{Chen26-ICCIP}.
Localized reconstruction frameworks bypass heavy network retraining by allowing the UE to rebuild a sparse linear codebook using fresh CSI samples gathered over a short post-shift interval~\cite{Joo26-TCOM}.
Recent research has introduced physics-based distribution alignment to bypass the need for target-domain data entirely, enabling zero-shot generalization in completely unseen environments~\cite{Wang26-TWC}.

Despite their effectiveness, most of these approaches share a common requirement: access to target-domain data, whether through online modification phases, fine-tuning procedures, or post-shift observation windows.
Because downlink channel estimation is executed exclusively at the UE, post-shift target-domain CSI is natively available only at the receiver.
Unless fresh target-domain parameters are explicitly fed back to the BS, the network-side decoder cannot re-align to the shifted environment.
True zero-shot generalization---maintaining high fidelity in unseen environments without any retraining or adaptation windows---remains a key challenge for DL-based CSI compression.

\subsection{Spatially-Scalable CSI Feedback}
Real-world BS deployments utilize diverse antenna configurations, such as $1 \times 32$ uniform linear arrays (ULAs) or $8 \times 8$ and $16 \times 4$ uniform planar arrays (UPAs).
However, many existing CSI compression studies lack spatial scalability, as the array shape is embedded in the CSI data dimensions, making pretrained models unusable under new array configurations.
Existing partial solutions include antenna subset selection, dimension-agnostic pooling, and modular per-antenna encoders~\cite{Zhong20-Sensors, He22-GCWkshps}.
Recently, coordinate-aware foundation models that map antenna indices to physical 3D coordinates have enabled zero-shot scale extrapolation~\cite{Zhang26-arXiv}.

While these approaches handle dynamic configurations, they still couple the downstream network's processing complexity, token length, or operational floating-point operations (FLOPs) to the underlying array dimensions.
A model designed for a $4 \times 4$ array, for instance, incurs substantially different computational costs when deployed on a $16 \times 8$ array.
Developing a truly architecture-independent framework with constant parameter counts and computational costs across varying antenna array sizes remains an open challenge for scalable CSI feedback.

\section{System Model}

We consider a single-cell FDD massive MIMO system, wherein a BS serves single-antenna UEs via orthogonal frequency division multiplexing (OFDM).
The BS is equipped with an $N_\text{r} \times N_\text{c}$ UPA, where $N_\text{r}$ and $N_\text{c}$ denote the number of antenna rows and columns, respectively.
Each element may support $P$ polarizations, yielding a total of $N_\text{ant} = P \, N_\text{r} N_\text{c}$ transmit antenna ports.
Throughout this work, we primarily consider the dual-polarized configuration ($P=2$), which is standard in 3GPP deployments, though single-polarized ($P=1$) datasets are also evaluated.
The total available operational bandwidth is divided into $N_\text{sc}$ OFDM subcarriers.
In this FDD architecture, uplink and downlink transmissions operate on distinct frequency bands, precluding the exploitation of channel reciprocity.
To enable spatial multiplexing and precoding at the BS, the UE must estimate the downlink CFR across all $N_\text{ant}$ antennas and $N_\text{sc}$ subcarriers, compress this high-dimensional matrix, and feed it back to the BS over a rate-limited uplink control channel.

The downlink CFR for the $n$-th subcarrier ($n = 0, 1, \dots, N_\text{sc}-1$) and the $m$-th antenna element can be modeled as a superposition of $L$ distinct multipath propagation components.
Specifically, the complex channel gain $\mathbf{H}[n, m] \in \mathbb{C}$ is formulated as:
\begin{equation}
    \mathbf{H}[n, m] = \sum_{l=1}^{L} \alpha_{l, m} e^{-j 2\pi f_n \tau_l},
\end{equation}
where $\alpha_{l, m} \in \mathbb{C}$ denotes the complex spatial amplitude of the $l$-th path as observed at the $m$-th antenna element, $\tau_l$ is the physical propagation delay of the $l$-th path, and $f_n$ is the frequency of the $n$-th subcarrier.
Stacking these responses across all subcarriers and antennas yields the full CFR matrix $\mathbf{H} \in \mathbb{C}^{N_\text{sc} \times N_\text{ant}}$.
A key physical property of this formulation is that while the complex gains $\{\alpha_{l, m}\}$ vary across antenna elements due to array geometry and polarization, the propagation delays $\{\tau_l\}$ are similar across all antennas.
This structural characteristic indicates that the CFR matrix $\mathbf{H}$ exhibits joint sparsity: it is intrinsically sparse in the delay domain but dense in the spatial domain.

In this work, we assume the UE is located in the far-field (Fraunhofer) region of the base station array.
Under this standard planar wavefront assumption, the physical propagation delay for a given multipath cluster is shared uniformly across all antenna elements (i.e., joint-spatial sparsity).
While the wavefront arrives at individual elements at marginally different absolute times, this microscopic difference is entirely absorbed as a linear spatial phase shift (Angle of Arrival) by the local complex amplitude $\alpha_{m}$ at each antenna.
Consequently, the dictionary matrix consisting of possible delays is constructed purely in the frequency-delay domain, allowing the subsequent LS solver to evaluate the spatial dimensions independently.

By applying an IFFT along the frequency axis of the CFR, the channel can be represented in the delay domain as the channel impulse response (CIR).
Because the channel energy is concentrated at the discrete physical propagation delays $\{\tau_l\}$, the resulting CIR is highly sparse.
The number of significant multipath taps, denoted as $K$, is typically much smaller than the total number of subcarriers ($K \ll N_\text{sc}$).
This delay-domain sparsity can be leveraged to reduce the amount of information that needs to be fed back to the BS~\cite{Gao15-TSP}.

Under typical UE mobility scenarios, the macroscopic parameters of the propagation environment, especially the delay profile, evolve smoothly over consecutive OFDM slots~\cite{Gao15-TSP}.
While the instantaneous complex phases of the spatial amplitudes fluctuate rapidly due to small-scale fading, the underlying physical delays $\{\tau_l\}$ exhibit strong temporal correlation.
This enables the use of a temporal window comprising $W$ previously estimated CFR snapshots to provide prior information regarding the delay support of the current channel realization.

In FDD massive MIMO systems, the BS periodically transmits orthogonal Channel State Information Reference Signals (CSI-RS) to facilitate downlink channel estimation at the UE~\cite{3gpp.38.214}.
Let $\mathbf{X}_p \in \mathbb{C}^{N_\text{ant} \times \tau_p}$ denote the orthogonal pilot matrix transmitted over $\tau_p$ symbol durations (where $\tau_p \ge N_\text{ant}$), designed such that $\mathbf{X}_p \mathbf{X}_p^H = P \mathbf{I}_{N_\text{ant}}$, where $P$ is the pilot transmission power.
At the $n$-th subcarrier, the received signal block $\mathbf{y}_n^T \in \mathbb{C}^{1 \times \tau_p}$ at a single-antenna UE is expressed as:
\begin{equation}
\mathbf{y}_n^T = \mathbf{h}_n^T \mathbf{X}_p + \mathbf{z}_n^T,\label{eq:received_pilot}
\end{equation}
where $\mathbf{h}_n \in \mathbb{C}^{N_\text{ant} \times 1}$ is the true downlink channel vector for subcarrier $n$, and $\mathbf{z}_n \sim \mathcal{CN}(\mathbf{0}, \sigma_z^2 \mathbf{I}_{\tau_p})$ represents the additive white Gaussian noise (AWGN) at the receiver.
To decouple the spatial channel components, the UE applies a standard LS estimator by correlating the received signal with the known pilot matrix.
The estimated channel vector for the $n$-th subcarrier is obtained as:
\begin{equation}
\widehat{\mathbf{h}}_n^T = \frac{1}{P} \mathbf{y}_n^T \mathbf{X}_p^H = \mathbf{h}_n^T + \widetilde{\mathbf{z}}_n^T,\label{eq:ls_estimation}
\end{equation}
where $\widetilde{\mathbf{z}}_n^T = \frac{1}{P} \mathbf{z}_n^T \mathbf{X}_p^H$ is the effective estimation noise.
By stacking the estimated vectors across all $N_\text{sc}$ subcarriers, we obtain the full observed CFR matrix fed into the compression framework:
\begin{equation}
\mathbf{H}_\text{obs} = \mathbf{H} + \mathbf{N},
\label{eq:noisy_cfr}
\end{equation}
where $\mathbf{H} \in \mathbb{C}^{N\text{sc} \times N_\text{ant}}$ is the ground-truth CFR matrix, and $\mathbf{N} \in \mathbb{C}^{N_\text{sc} \times N_\text{ant}}$ is the aggregated equivalent estimation noise matrix.
By aggregating the effective estimation noise vectors $\widetilde{\mathbf{z}}_n^T$, the equivalent estimation noise matrix can be explicitly expressed in terms of the received over-the-air noise and the pilot matrix as:
\begin{equation}\mathbf{N} = \frac{1}{P} \mathbf{Z} \mathbf{X}p^H,
\label{eq:matrix_noise}
\end{equation}
where $\mathbf{Z} \in \mathbb{C}^{N\text{sc} \times \tau_p}$ is the aggregate AWGN matrix at the receiver, whose $n$-th row corresponds to the noise vector $\mathbf{z}_n^T$ experienced at subcarrier $n$.
Because the pilot matrix $\mathbf{X}_p$ is orthogonal, the linear transformation $\frac{1}{P}(\cdot)\mathbf{X}_p^H$ projects the raw receiver noise into the spatial antenna domain, directly dictating the input CSI-RS signal-to-noise ratio (SNR).
This SNR, defined as $\text{SNR}_\text{CSI-RS} = \Vert{}\mathbf{H}\Vert{}_F^2 / \Vert{}\mathbf{N}\Vert{}_F^2$, determines the quality of the initial channel estimate available to the compressor.
In the ideal case where estimation noise is completely absent, the compressor operates directly on the true physical channel $\mathbf{H}$.

The fundamental objective of the CSI feedback problem is to design an efficient compression and reconstruction pipeline.
Let the UE possess the estimated downlink CFR matrix $\mathbf{H}$ (or $\mathbf{H}_\text{obs}$ under noisy conditions) along with a historical window of previous CFRs.
The UE utilizes an encoder $E(\cdot)$ to map the full-dimensional CFR into a compressed latent representation $\mathbf{z} = E(\mathbf{H})$, where the dimension of $\mathbf{z}$, denoted as $M$, satisfies $M \ll P N_\text{sc} N_\text{ant}$.
The compressed vector $\mathbf{z}$ is then transmitted over the uplink channel.
At the BS, a decoder $D(\cdot)$ reconstructs the approximated CFR $\mathbf{\hat{H}} = D(\mathbf{z})$.
The efficacy of the proposed feedback system is evaluated by $\text{CR} = \frac{M}{P N_\text{sc} N_\text{ant}}$, and the frequency-domain CFR Normalized Mean Square Error (NMSE), or CFR-NMSE in short.
CFR-NMSE is defined as
\begin{equation}
\text{NMSE} = 10 \log_{10} \left( \frac{|\mathbf{H} - \mathbf{\hat{H}}|F^2}{|\mathbf{H}|F^2} \right).
\label{eq:cfr_nmse}
\end{equation}
An important distinction arises between this metric and those utilized in recent literature.
The majority of prior DL-based works operate on a truncated delay-angular domain patch extracted from the full channel matrix, reporting the NMSE computed solely over this cropped representation (hereafter, patch NMSE).
Because truncation inherently discards signal energy outside the selected patch, patch NMSE systematically understates the true reconstruction error.
In contrast, our adopted CFR-NMSE quantifies the reconstruction fidelity over the complete CFR matrix $\mathbf{H} \in \mathbb{C}^{N\text{sc} \times N\text{ant}}$, providing a much more comprehensive measure of compression fidelity.
Therefore, the overarching goal of our framework is to design an encoding and decoding architecture that minimizes the full CFR-NMSE subject to a target CR, while ensuring the encoding process remains computationally tractable for the UE.

\section{OMP-based CIR Recovery}

The standard OMP algorithm recovers the sparse CIR by iteratively building a support set of the most dominant propagation delays.
At each iteration, the algorithm correlates the current residual signal with a predefined discrete delay dictionary, selects the dictionary atom that exhibits the highest correlation energy, and orthogonalizes the residual by performing a LS projection of the received CFR onto the active support set.
As outlined in Algorithm~\ref{alg:vanilla_omp}, this greedy pursuit process repeats until $K$ sparse taps are extracted or the residual energy falls below a predetermined threshold.

\begin{algorithm}[t]
\caption{OMP Algorithm for CIR Estimation}
\label{alg:vanilla_omp}
\begin{algorithmic}[1]
\Require Received CFR $\mathbf{h}$, Dictionary $\mathbf{\Phi}$, Max paths $K$, Threshold $\epsilon$, Oversampling factor $S$
\State $\mathbf{r}_0 \gets \mathbf{h}$ \Comment{Initialize residual}
\State $\Lambda_0 \gets \emptyset$ \Comment{Initialize set of selected indices}
\State $i \gets 1$
\While{$i \leq K$ and $||\mathbf{r}_{i-1}||_2^2 > \epsilon$}
    \State $\mathbf{c} \gets S N_\text{sc} \cdot \text{IFFT}(\bar{\mathbf{r}}_{i-1})$ \Comment{Calculate correlation}
    \State $k \gets \arg\max_j \sum_{n=1}^{N_\text{sym}} |\mathbf{c}_{j,n}|^2$ \Comment{Find atom with maximum energy}
    \State $\Lambda_i \gets \Lambda_{i-1} \cup \{k\}$ \Comment{Update support set}
    \State $\mathbf{\Phi}_i \gets \mathbf{\Phi}[:, \Lambda_i]$ \Comment{Construct active dictionary}
    \State $\mathbf{x}_i \gets (\mathbf{\Phi}_i^H \mathbf{\Phi}_i + \lambda \mathbf{I})^{-1} \mathbf{\Phi}_i^H \mathbf{h}$ \Comment{Least squares}
    \State $\mathbf{\hat{h}}_i \gets \mathbf{\Phi}_i \mathbf{x}_i$ \Comment{Projected signal}
    \State $\mathbf{r}_i \gets \mathbf{h} - \mathbf{\hat{h}}_i$ \Comment{Update orthogonal residual}
    \State $i \gets i + 1$
\EndWhile
\State \textbf{return} Sparse gains $\mathbf{x}$ and selected delays $\Lambda$
\end{algorithmic}
\end{algorithm}

\begin{figure}[!t]
    \centering
    \includegraphics[width=\if\numcol1 0.6 \else 0.85\fi\columnwidth]{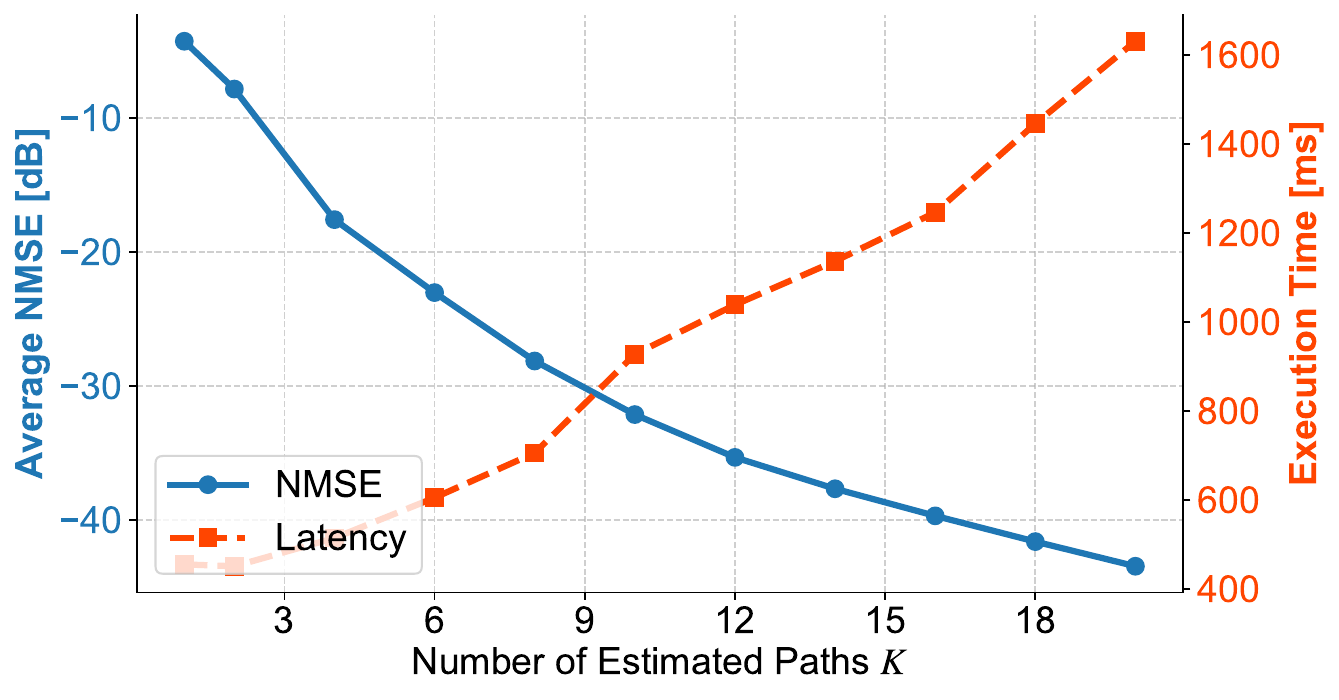}
    \caption{NMSE and latency of OMP-based CSI compression under different number of estimated taps.}
    \label{fig:omp_nmse_latency1}
\end{figure}

Despite its accuracy, the computational complexity of the iterative OMP procedure scales prohibitively with the number of estimated paths~\cite{Du24-ICSPCC, Oyerinde23-ICSPCS}.
At the $i$-th iteration, the algorithm requires evaluating the correlation across the entire oversampled dictionary, typically implemented via an $N_\text{sc}$-point IFFT with cost $\mathcal{O}(S N_\text{sc} \log(S N_\text{sc}))$, where $S$ is the oversampling factor.
The orthogonalization step requires solving an increasingly large system of equations, demanding $\mathcal{O}(i^2 N_\text{sc})$ floating-point operations.
For massive MIMO systems requiring the estimation of tens of multipath components, this compounded complexity becomes prohibitive.
As shown in Fig.~\ref{fig:omp_nmse_latency1}, executing standard OMP to recover $K=10$ paths requires approximately 1000~ms of processing time per feedback instance.
Notably, this latency was recorded within a high-performance simulation environment (AMD Ryzen Threadripper 7970X, dual NVIDIA RTX 6000 Ada GPUs, 256~GB RAM) utilizing heavily vectorized, multi-threaded CPU algebraic backends.
Even with this computational advantage, OMP exceeds the sub-millisecond latency budget of 5G NR by over three orders of magnitude.
This indicates that execution on power- and memory-constrained UE hardware would be completely prohibitive.

\subsection{Challenges of One-Shot OMP}
A computationally cheaper alternative to this iterative pursuit would be a ``one-shot'' approach: performing a single IFFT on the CFR and selecting the $K$ highest magnitude peaks in the resulting delay-domain signal.
However, this approach fails catastrophically due to two fundamental phenomena: spectral leakage and mutual coherence.

\subsubsection{Spectral Leakage Problem}
In any real-world digital communication system, the continuous physical CFR is only observed over a finite, strictly limited bandwidth $B$.
The physical receiver hardware naturally imposes a rectangular band-pass filter, mathematically represented as $\text{rect}(f/B)$, where $\text{rect}(\cdot)$ equals 1 for $[-1/2, 1/2]$ and 0 elsewhere.
According to the convolution theorem of Fourier analysis, multiplication by a rectangular window in the frequency domain translates directly to convolution with a sinc function in the delay domain.

Consequently, a pristine physical multipath impulse (e.g., a Dirac delta function representing a pure reflection) is smeared into a heavily rippled, multi-tap discrete signal characterized by a broad main lobe and infinitely decaying side lobes.
When a naive one-shot algorithm observes this filtered delay-domain profile, it lacks the mathematical mechanism to reliably distinguish between the genuine energy of a weaker, secondary physical path and the massive sinc-sidelobe generated by a highly dominant primary path.

The traditional, iterative OMP algorithm natively resolves this ambiguity.
Because the main lobe and side lobes are mathematically coupled, once iterative OMP identifies the strongest peak and computes its complex gain, the subsequent orthogonal projection step precisely subtracts the reconstructed frequency-domain signature of that path from the observed CFR.
This inherently subtracts the exact mathematical cause of the time-domain sidelobes, leaving the newly updated residual clean of the primary tap's interference.
A one-shot approach, lacking this sequential interference cancellation, suffers from massive false-alarm rates, repeatedly and incorrectly identifying sidelobe ripples as active channel support.

\subsubsection{Non-Orthogonal Atoms and Grid Mismatch}
The second critical failure point for discrete one-shot estimation arises from the necessary discretization of the continuous physical delay space, known as grid mismatch~\cite{Pali22-SPL}.
Standard IFFT operations implicitly assume that true multipath delays align perfectly with the integer sampling bins of the discrete digital grid, spaced at intervals of $1/(S N_\text{sc} \Delta f)$, where $\Delta f$ represents the subcarrier spacing.
In physical reality, propagation delays take arbitrary, continuous fractional values.
When a physical path delay falls between these bins, its energy bleeds into multiple adjacent grid points, destroying the fundamental sparsity assumption.

To mitigate grid mismatch and capture these fractional delays with higher fidelity, the dictionary matrix $\mathbf{\Phi}$ is heavily oversampled by a factor $S > 1$.
However, oversampling fundamentally destroys the linear orthogonality between adjacent dictionary atoms.
Assuming a system utilizing $N_\text{sc}$ subcarriers with a subcarrier spacing $\Delta f$, the inner product between two dictionary atoms $\mathbf{a}(\tau_1)$ and $\mathbf{a}(\tau_2)$ representing arbitrary delays $\tau_1$ and $\tau_2$ is strictly dictated by the Dirichlet kernel:
\begin{align}
\langle \mathbf{a}(\tau_1), \mathbf{a}(\tau_2) \rangle &= \sum_{k=0}^{N_\text{sc}-1} e^{-j 2\pi k \Delta f \tau_1} (e^{-j 2\pi k \Delta f \tau_2})^* \\
&= \frac{1 - e^{-j 2\pi N_\text{sc} \Delta f \Delta \tau}}{1 - e^{-j 2\pi \Delta f \Delta \tau}}.
\end{align}

For perfect orthogonality ($\langle \mathbf{a}(\tau_1), \mathbf{a}(\tau_2) \rangle = 0$), the physical delays must be separated by an exact integer multiple of the inverse system bandwidth $B = N_\text{sc} \Delta f$, such that $\Delta\tau = m/B$ for any integer $m$.
Because oversampling actively forces adjacent atoms to be separated by a distance much less than $1/B$, the entire dictionary becomes highly coherent.
In a one-shot estimation framework, the energy from a single true fractional path bleeds heavily into multiple adjacent, non-orthogonal atoms.
A parallel peak-selection protocol will mistakenly identify this dense cluster of highly correlated atoms as multiple distinct physical paths, destroying the CR and generating an erroneous support set.

While various off-grid optimization strategies such as alternating gradient descent, Newton refinement, or atomic norm soft thresholding have been proposed to iteratively push grid points toward the true fractional delays, these techniques introduce severe non-linear computational complexity that violates the stringent low-latency constraints of standard baseband processing.

\subsection{Error Bound of Imperfect Delay Estimation}
\label{subsec:delay_error_bound}
Here, we study how tap delay errors impact estimation performance.
We also analyze the ability of LS estimation to mitigate tap estimation errors in wideband OFDM systems.
We show that LS estimation cannot fully compensate for minor inaccuracies in estimated tap delays when calculating complex amplitudes.

A complex LS amplitude estimate, denoted as $\hat{a} = |\hat{a}|e^{j\theta}$, applies a constant phase rotation ($\theta$) uniformly across all subcarriers.
In contrast, a physical delay error $\Delta \tau$ introduces a frequency-dependent phase rotation of $-2\pi f_k \Delta \tau$.
This manifests as a linear phase ramp across the frequency band.
Consequently, the LS estimator can only perfectly align the phase at the center frequency ($f=0$).
At the band edges, the phase mismatch grows linearly.
The LS solver is mathematically incapable of altering the slope of the phase across subcarriers; it can only shift the intercept.

Here, we rigorously bound the CSI reconstruction error caused by this phenomenon to demonstrate the necessity of high-precision delay extraction.

\subsubsection{Reconstruction Error Bound for a Single Tap}
Let $N_\text{sc}$ denote the number of subcarrier frequencies, denoted as $f_k$.
Assume the true continuous CIR consists of a single tap with complex amplitude $a$ and delay $\tau$.
The true CFR vector $\mathbf{h} \in \mathbb{C}^{N_\text{sc}}$ is defined component-wise as:
\begin{equation}
    [\mathbf{h}]_k = a e^{-j 2\pi f_k \tau}.
\end{equation}

Let the estimated delay be $\hat{\tau} = \tau + \Delta \tau$.
The dictionary matrix $\mathbf{\hat{A}}$ for this single-tap estimation reduces to a single column vector $\mathbf{\hat{v}}$, where:
\begin{equation}
    [\mathbf{\hat{v}}]_k = e^{-j 2\pi f_k (\tau + \Delta \tau)}.
\end{equation}

The LS reconstructed signal $\mathbf{\hat{h}}$ is the orthogonal projection of the true signal $\mathbf{h}$ onto the estimated basis vector $\mathbf{\hat{v}}$:
\begin{equation}
    \mathbf{\hat{h}} = \mathbf{\hat{v}} (\mathbf{\hat{v}}^H \mathbf{\hat{v}})^{-1} \mathbf{\hat{v}}^H \mathbf{h} = \mathbf{P}_{\mathbf{\hat{v}}} \mathbf{h},
\end{equation}
where $\mathbf{P}_{\mathbf{\hat{v}}}$ is the orthogonal projection matrix.
We define the residual error vector as $\mathbf{e} = \mathbf{h} - \mathbf{\hat{h}} = (\mathbf{I} - \mathbf{P}_{\mathbf{\hat{v}}}) \mathbf{h}$.

Noting that the true signal can be factored as $\mathbf{h} = a \mathbf{v}$ (where $\mathbf{v}$ is the true basis vector), we can express $\mathbf{v}$ in terms of the estimated vector and the deviation between them: $\mathbf{v} = \mathbf{\hat{v}} - \Delta \mathbf{v}$.
Substituting this into the error equation yields:
\begin{equation}
    \mathbf{e} = a (\mathbf{I} - \mathbf{P}_{\mathbf{\hat{v}}}) (\mathbf{\hat{v}} - \Delta \mathbf{v}).
\end{equation}

Since the projection operator annihilates $\mathbf{\hat{v}}$, this simplifies to $\mathbf{e} = -a (\mathbf{I} - \mathbf{P}_{\mathbf{\hat{v}}}) \Delta \mathbf{v}$.
Taking the $\ell_2$-norm, and using the property of orthogonal projectors that $\Vert{}\mathbf{I} - \mathbf{P}_{\mathbf{\hat{v}}}\Vert{}_2 \leq 1$, we obtain:
\begin{equation}
\|\mathbf{e}\|_2 \leq |a| |\Delta \mathbf{v}|_2.
\end{equation}

Next, we evaluate the geometric distance between the true and estimated basis vectors:
\begin{align}
\|\Delta \mathbf{v}\|_2^2
&= \sum_{k=1}^{N_\text{sc}} \left| e^{-j 2\pi f_k \tau} - e^{-j 2\pi f_k (\tau + \Delta \tau)} \right|^2 \\
&= \sum_{k=1}^{N_\text{sc}} \left| 1 - e^{-j 2\pi f_k \Delta \tau} \right|^2 \\
&\leq \sum_{k=1}^{N_\text{sc}} (-2\pi f_k \Delta \tau)^2 \\
&= 4\pi^2 \Delta \tau^2 \sum_{k=1}^{N_\text{sc}} f_k^2.
\end{align}
where we factored out the true phase term $e^{-j 2\pi f_k \tau}$ in the second equality.
The trigonometric identity $|1 - e^{jx}|^2 = 4 \sin^2(x/2)$ and the inequality $4 \sin^2(x/2) \leq x^2$ for $x \in \mathbb{R}$, is then used to bound the distance.

Defining the frequency spread energy as a constant $E_f = \sum_{k=1}^{N_\text{sc}} f_k^2$, we establish the absolute error energy bound:
\begin{equation}
    \|\mathbf{e}\|_2^2 \leq 4\pi^2 E_f |a|^2 \Delta \tau^2.
\end{equation}

Given that the total signal energy is $\|\mathbf{h}\|_2^2 = N_\text{sc} |a|^2$, the CFR-NMSE for a single tap is bounded by:
\begin{equation}
    \text{CFR-NMSE} \leq \frac{4\pi^2 E_f}{N_\text{sc}} \Delta \tau^2.
\end{equation}
This confirms that the CFR-NMSE grows quadratically with respect to the delay estimation error $\Delta \tau$, proving that LS regression is mathematically incapable of mitigating this error~\cite{Xi20-TSP}.

\subsubsection{Generalization to Multiple Taps}
We generalize this bound for a true channel characterized by $L$ taps, such that $\mathbf{h} = \mathbf{A} \mathbf{a}$, where $\mathbf{A} = [\mathbf{v}_1, \dots, \mathbf{v}_L]$ and $\mathbf{a} \in \mathbb{C}^L$.
Let the estimated delays be $\hat{\tau}_i = \tau_i + \Delta \tau_i$, which form the estimated dictionary $\mathbf{\hat{A}}$.

The multi-tap error vector is $\mathbf{e} = (\mathbf{I} - \mathbf{P}_{\mathbf{\hat{A}}}) \mathbf{A} \mathbf{a}$.
Expressing the true dictionary as $\mathbf{A} = \mathbf{\hat{A}} - \Delta \mathbf{A}$, where the $i$-th column of $\Delta \mathbf{A}$ is $\Delta \mathbf{v}_i$, we find:
\begin{equation}
    \mathbf{e} = -(\mathbf{I} - \mathbf{P}_{\mathbf{\hat{A}}}) \Delta \mathbf{A} \mathbf{a}.
\end{equation}

Taking the norm and applying the triangle inequality yields:
\begin{equation}
    \|\mathbf{e}\|_2 \leq \|\Delta \mathbf{A} \mathbf{a}\|_2 \leq \sum_{i=1}^L |a_i| \|\Delta \mathbf{v}_i\|_2.
\end{equation}

Substituting the single-column bound derived previously ($\|\Delta \mathbf{v}_i\|_2 \leq 2\pi |\Delta \tau_i| \sqrt{E_f}$), we establish the final multi-tap error energy bound:
\begin{equation}
    \|\mathbf{e}\|_2^2 \leq 4\pi^2 E_f \left( \sum_{i=1}^L |a_i| |\Delta \tau_i| \right)^2.
\end{equation}

\subsubsection{Implications for CSI Reconstruction}
This derivation highlights two critical properties governing wideband sparse channel estimation:
\begin{itemize}
    \item \textbf{Bandwidth Multiplier:} The error bound is linearly scaled by the frequency spread energy $E_f = \sum_{k=1}^{N_\text{sc}} f_k^2$.
Consequently, as the OFDM bandwidth widens, the inability of the LS solver to compensate for $\Delta \tau$ becomes exponentially more destructive.
    \item \textbf{Amplitude Weighting:} The reconstruction error is weighted by the true tap amplitudes $|a_i|$.
A minuscule delay error $\Delta \tau$ on a dominant line-of-sight (LoS) path will severely degrade the overall CFR-NMSE, regardless of the precision of the complex coefficient optimization.
\end{itemize}

These fundamental limitations, namely prohibitive iterative complexity, vulnerability to spectral leakage, discrete dictionary mismatch, and the extreme sensitivity of the reconstruction error to delay precision, motivate the development of a neural one-shot estimation framework.
A neural network can be trained on ray-traced data to denoise sidelobes and disambiguate closely spaced paths from temporal CFR sequences.
Coupling such a network with a differentiable sub-grid extraction mechanism enables prediction of continuous, fractional delay positions in a single forward pass, bypassing both the iterative loops of classical algorithms and the resolution limits of discrete dictionaries.

\section{Temporal Channel Estimation and Feedback}

\begin{figure*}[!t]
    \centering
    \includegraphics[width=\if\numcol1 1.0 \else 1.0\fi\linewidth]{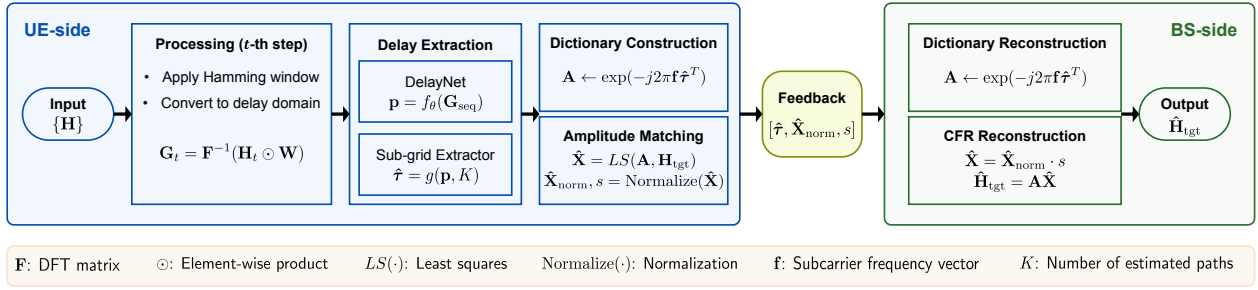}
    \caption{Pipeline of TAP.}
    \label{fig:pipeline}
\end{figure*}

To address the limitations of classical iterative pursuit and rigid neural architectures, we propose TAP.
As depicted in Fig.~\ref{fig:pipeline} and formalized in Algorithm~\ref{alg:st_omp}, TAP is designed as a fully differentiable, one-shot computational pipeline comprising three cascaded stages: (i) temporal delay prediction, (ii) differentiable sub-grid extraction, and (iii) spatial amplitude estimation.
By ensuring every stage permits backpropagation, the entire framework seamlessly bridges structural sparse recovery with deep learning, enabling end-to-end training using composite loss functions.

\begin{algorithm}[t]
\caption{TAP Pipeline for MIMO CSI Compression}
\label{alg:st_omp}
\begin{algorithmic}[1]
\Require Previous MIMO CFRs $\mathbf{H}_\text{prev}$, Current (Target) CFR $\mathbf{H}_\text{tgt}$, Max paths $K$, Regularization $\lambda$
\Ensure Reconstructed Channel Frequency Response $\mathbf{\hat{H}}$

\Statex \textbf{UE Side (Estimation, Compression \& Feedback):}
\Require DelayNet $f_\theta$, Delay Extractor $g$
\State $\mathbf{H}_\text{seq} \gets [\mathbf{H}_\text{prev}, \mathbf{H}_\text{tgt}]$
\State $\mathbf{G}_\text{seq} \gets \text{IFFT}(\mathbf{H}_\text{seq})$ 
\State $\mathbf{p} \gets f_\theta(\mathbf{G}_\text{seq})$ \Comment{Generate delay heatmap}
\State $\boldsymbol{\hat{\tau}} \gets g(\mathbf{p}, K)$ \Comment{Extract delays}
\State $\mathbf{A} \gets \exp(-j 2\pi \mathbf{f} \boldsymbol{\hat{\tau}}^T)$ \Comment{Construct dictionary}
\State $\mathbf{\hat{X}} \gets (\mathbf{A}^H \mathbf{A} + \lambda \mathbf{I})^{-1} \mathbf{A}^H \mathbf{H}_\text{tgt}$ \Comment{Compute amplitudes}
\State $\mathbf{\hat{X}}_\text{norm}, s \gets \text{Normalize}(\mathbf{\hat{X}})$
\State \textbf{return} Delays $\boldsymbol{\hat{\tau}}$, Amplitudes $\mathbf{\hat{X}}_\text{norm}$, and Scale $s$

\Statex \textbf{BS Side (Channel Reconstruction):}
\State $\mathbf{\hat{X}} \gets s \cdot \mathbf{\hat{X}}_\text{norm}$ \Comment{Restore physical amplitude scaling}
\State $\mathbf{A} \gets \exp(-j 2\pi \mathbf{f} \boldsymbol{\hat{\tau}}^T)$ \Comment{Reconstruct dictionary matrix}
\State $\mathbf{\hat{H}} \gets \mathbf{A} \mathbf{\hat{X}}$ \Comment{Reconstruct full MIMO CFR}
\State \textbf{return} Reconstructed CFR $\mathbf{\hat{H}}$
\end{algorithmic}
\end{algorithm}

\subsection{Temporal Input Representation}
The input to the system leverages temporal channel coherence.
Let $\mathbf{H}_\text{seq} = [\mathbf{H}_{t-W+1}, \dots, \mathbf{H}_t] \in \mathbb{C}^{W \times P \times N_\text{r} \times N_\text{c} \times N_\text{sc}}$ represent a temporally windowed sequence of length $W$, encompassing the current and $W-1$ historical CFR snapshots.
Rather than processing the raw frequency-domain data, TAP applies a Hamming-windowed IFFT across the subcarrier axis at each time step $t$ to produce a delay-domain tensor $\mathbf{G}_t$.
These are then concatenated to transform the sequence into the spatial-delay domain tensor $\mathbf{G}_\text{seq} \in \mathbb{C}^{W \times P \times N_\text{r} \times N_\text{c} \times N_\text{sc}}$.
This representation addresses the spectral leakage problem identified in the previous section: the Hamming window suppresses sinc-sidelobes, effectively eliminating phantom peaks.
Stacking the sequence along the temporal dimension provides motion-based disambiguation cues, allowing the neural network to distinguish consistent, physically moving propagation paths from transient noise artifacts.

\subsection{DelayNet Architecture}
The core of the delay prediction engine is DelayNet, a 1D convolutional residual network that maps the high-dimensional spatial-delay-temporal input into a 1D probability heatmap over the discrete delay axis.
DelayNet employs depthwise separable convolutions combined with an initial spatial-averaging layer.
This design decouples the network's parameters from the spatial dimensions of the input; the convolution weights do not scale with the number of BS antennas.
The architecture utilizes grouped residual blocks with GroupNorm and GELU activations to ensure stable training dynamics independent of batch size.
A sigmoid activation at the output layer produces a multi-label probability heatmap $\mathbf{p} \in [0, 1]^{N_\text{d}}$, permitting the simultaneous detection of multiple dominant paths that share closely spaced delay bins.
Here, $N_\text{d}$ denotes the estimated heatmap length.
The antenna-averaging and depthwise convolutional structures ensure that the DelayNet parameter count remains constant regardless of the $N_\text{r} \times N_\text{c}$ antenna geometry deployed at the base station.

\subsection{Differentiable Sub-Grid Delay Extraction}
To overcome the discrete dictionary mismatch problem without resorting to computationally expensive dictionary oversampling, TAP utilizes a differentiable parabolic interpolation operator, denoted as $g(\mathbf{p}, K)$, to extract $K$ continuous, sub-grid delay values $\boldsymbol{\hat{\tau}} = [\hat{\tau}_1, \dots, \hat{\tau}_K]^T$ from the discrete output heatmap $\mathbf{p}$.

For each of the top $K$ local peaks identified at integer bin index $i$, the corresponding fractional sub-bin offset $\Delta i \in [-0.5, 0.5]$ is computed using the probability values of the peak $p_i$ and its immediate neighbors $p_{i-1}$ and $p_{i+1}$.
By fitting a parabola to these three points, the true vertex offset is analytically extracted as:
\begin{equation}
    \Delta i = \frac{1}{2} \frac{p_{i-1} - p_{i+1}}{p_{i-1} - 2 p_i + p_{i+1}}.
\end{equation}
The continuous physical delay is then obtained by scaling the fractional index by the time-domain sampling interval $\Delta t$, such that $\hat{\tau}_k = (i + \Delta i) \Delta t$.

This mathematically continuous extraction allows the estimated delays to fall precisely between rigid grid points, successfully resolving off-grid multipath components.
More importantly, this operator is fully differentiable with respect to the input probabilities.
This differentiability provides a transparent conduit for gradient flow, enabling the reconstruction error gradient originating from the downstream solver to backpropagate directly into DelayNet.

\subsection{Spatial Amplitude Estimation}
Given the extracted continuous delays $\boldsymbol{\hat{\tau}}$, the spatial amplitudes across all $N_\text{ant}$ antenna elements are recovered simultaneously using a standard regularized least squares solver, denoted as $LS()$.
A continuous Fourier dictionary matrix $\mathbf{A} \in \mathbb{C}^{N_\text{sc} \times K}$ is constructed exactly as $\mathbf{A}[n, k] = \exp(-j 2\pi f_n \hat{\tau}_k)$.
Exploiting the joint-spatial sparsity of the CFR, we solve for the complex spatial amplitudes $\mathbf{\hat{X}} \in \mathbb{C}^{K \times N_\text{ant}}$:
\begin{equation}
    \mathbf{\hat{X}} = (\mathbf{A}^H \mathbf{A} + \lambda \mathbf{I})^{-1} \mathbf{A}^H \mathbf{H}_\text{tgt}.
\end{equation}
Because the continuous basis $\mathbf{A}$ is identical across all antennas, the generalized inverse only needs to be computed once per UE.
While this solver is mathematically identical to the orthogonalization step in classical iterative OMP, executing it only once after a single-shot neural delay prediction dramatically reduces the computational burden.
This operation consumes roughly 18~MFLOPs, providing exact spatial recovery without introducing any learned parameters.

\subsection{Decoder-Free BS Reconstruction}
Upon receiving the compressed feedback payload, the BS reconstructs the full-dimensional downlink CFR matrix.
Crucially, unlike deep learning-based CSI feedback schemes that require deploying and executing a computationally heavy neural decoder at the base station, TAP offers a completely decoder-free reconstruction architecture.
Because the feedback payload explicitly represents the physical channel parameters, the BS reconstructs the CFR simply by multiplying the spatial amplitudes by the Fourier dictionary matrix $\mathbf{A}$, effectively applying an inverse Fourier transform: $\mathbf{\hat{H}}_\text{tgt} = \mathbf{A} \mathbf{\hat{X}}$.
This purely algebraic reconstruction entirely eliminates the need for maintaining environment-specific neural weights at the network side, drastically reducing BS-side inference latency and deployment overhead.
\subsection{Composite Loss and End-to-End Training}
The entire TAP pipeline is trained end-to-end utilizing a composite objective function that combines localized delay supervision with global reconstruction constraints.
The loss is formulated as a weighted sum of a heatmap mean square error (MSE) loss, $\mathcal{L}_{\text{heatmap}}$, and a CFR-NMSE loss, $\mathcal{L}_{\text{CFR}}$.

To calculate $\mathcal{L}_{\text{heatmap}}$, a ground-truth probability heatmap $\mathbf{p}^\text{target} \in [0, 1]^{N_\text{d}}$ is dynamically generated during training using the true physical delays $\tau_l$ and their corresponding linear amplitudes $\alpha_l$ extracted from the ray-tracing simulator.
For each true path, a localized Gaussian distribution is placed at the exact fractional delay bin $\tau_l / \Delta t$.
The height of each Gaussian is scaled by the relative amplitude of the path, ensuring that the network prioritizes the most dominant propagation components.
Mathematically, the target heatmap is constructed as:
\begin{equation}
    p_i^\text{target} = \min \left( \sum_{l=1}^L \frac{|\alpha_l|}{\max_j |\alpha_j|} e^{\left( - \frac{(i - \tau_l / \Delta t)^2}{2 \sigma^2} \right)}, 1.0 \right),
\end{equation}
where $\sigma$ controls the width of the Gaussian targets.
The heatmap loss is then computed as $\mathcal{L}_{\text{heatmap}} = \|\mathbf{p} - \mathbf{p}^\text{target}\|_2^2$.

This dual-loss methodology accelerates convergence: $\mathcal{L}_{\text{heatmap}}$ provides explicit, aggressive early-stage supervision for the delay extraction mechanism, preventing the gradients from vanishing through the non-linear LS solver.
Simultaneously, the end-to-end $\mathcal{L}_{\text{CFR}}$ guarantees that the final spatial amplitudes and overall channel reconstruction are optimally tuned for maximum fidelity.

\section{Performance Evaluation}

\subsection{Simulation Setup}
We evaluate the proposed TAP framework across five diverse propagation environments to test its performance and generalizability.
Simulation settings are detailed in Table~\ref{tab:simulation_settings}.
We use evaluation parameters aligned with the 3GPP Frequency Range 1 (FR1) standard, employing a 100 MHz bandwidth~\cite{3gpp.38.101.1}.
The datasets encompass two Sionna RT simulator data---Urban Macrocell (UMa), and Indoor Factory (InF)---as well as two DeepMIMO datasets each representing Urban Microcell (UMi) and Indoor Hotspot (InH), and a 3GPP stochastic CDL-A channel model generated using Sionna-PHY~\cite{3gpp.38.901, hoydis23-arXiv_SionnaRT, hoydis23-arXiv_Sionna, Alkhateeb19-arXiv}.
These environments provide comprehensive coverage of indoor, outdoor, line-of-sight (LOS), and non-LOS conditions with varying delay spreads.


\begin{table}[t]
\centering
\caption{Simulation Settings for Evaluation Environments}
\label{tab:simulation_settings}
\adjustbox{width=\if\numcol1 0.6 \else 1.0 \fi\linewidth}{
\begin{tabular}{@{}lcccccc@{}}
\toprule
\multirow{2}{*}{\textbf{Parameter}} & \multicolumn{6}{c}{\textbf{Environments}} \\ \cmidrule(l){2-7} 
 & \textbf{UMa} & \textbf{InF} & \textbf{CDL-A} & \textbf{UMi} & \textbf{InH} & \textbf{ULA} \\ \midrule
UE velocity (m/s) & 16.6 & 1.0 & -- & -- & -- & -- \\
Downlink frequency $f_c$ (GHz) & 2.14 & 2.14 & 2.14 & 3.5 & 2.5 & 3.5 \\
Subcarrier spacing (kHz) & 60 & 60 & 60 & 60 & 60 & 30 \\
Number of subcarriers $N_\text{sc}$ & 1620 & 1620 & 1620 & 1620 & 1620 & 1024 \\
Number of BS antenna rows $N_\text{r}$ & 8 & 8 & 8 & 8 & 8 & 1 \\
Number of BS antenna columns $N_\text{c}$ & 8 & 8 & 8 & 8 & 8 & 32 \\
Antenna polarization $P$ & 2 & 2 & 2 & 1 & 1 & 1 \\
Oversample factor $S$ & 4 & 4 & 4 & 4 & 4 & -- \\
$T_{\text{max}}$ ($\mu$s) & 1.5 & 0.7 & 1.5 & 1.5 & 0.7 & -- \\
Temporal window length ($W$) & 9 & 9 & 1 & 9 & 9 & -- \\ \bottomrule
\end{tabular}
}
\end{table}

\subsection{Baseline Methods}
To benchmark the proposed TAP framework against the established state-of-the-art, we compare it against distinct baselines spanning four methodological categories: classical iterative CS algorithms (OMP, LASSO, and TVAL3), standardized codebook-based approaches (3GPP Type-1 and Type-2 codebooks)~\cite{3gpp.38.214}, and deep learning-based autoencoders (CsiNet-UPA and CsiNet-LSTM-UPA).

The original CsiNet architecture was strictly constrained to $1 \times 32$ ULAs~\cite{Wen18-WCL}.
To provide a fair deep learning baseline for modern massive MIMO systems, we expanded CsiNet to operate on UPAs by adapting its fully connected bottleneck and convolutional layers to support 2D spatial dimensions, which we denote as CsiNet-UPA.
Furthermore, to evaluate a learning-based baseline capable of exploiting temporal correlation, we similarly adapted the CsiNet-LSTM architecture to support UPA configurations, which we denote as CsiNet-LSTM-UPA~\cite{Wang19-WCL}.
To validate this expanded architecture and address potential skepticism regarding its effectiveness, we evaluate it on the original $1 \times 32$ ULA configuration.
The results are listed in Table~\ref{tab:ula_comparison}, along with the TAP's train dataset type.
As shown in the Table, our CsiNet architecture implementation achieves a mean CFR-NMSE of $-15.54$~dB, which aligns with the performance reported in the original literature.
We also note that TAP outperforms CsiNet architecture in terms of CFR-NMSE.
Specifically, UMa-trained TAP model achieves mean NMSE of $-25.88$~dB and InF-trained TAP model achieves $-37.01$~dB on a $1\times 32$ ULA architecture.
This substantial 10 to 21~dB performance gap is consistent even at the 90th and 95th percentile worst-case thresholds, highlighting the fundamental superiority of combining physical structural extraction with neural precision over dense fully connected autoencoders.

\begin{table}
\centering
\caption{NMSE Performance Comparison on $1 \times 32$ ULA Architecture}
\label{tab:ula_comparison}
\adjustbox{width=\if\numcol1 0.3 \else 0.6 \fi\linewidth}{
\begin{tabular}{@{}lccc@{}}
\toprule
\multirow{2}{*}{\textbf{Method}} & \multicolumn{3}{c}{\textbf{CFR NMSE {[}dB{]}}} \\ \cmidrule(l){2-4} 
 & $P_{90}$ & $P_{95}$ & Mean \\ \midrule
Proposed (UMa) & -22.33 & -21.68 & -25.88 \\
Proposed (InF) & -35.05 & -34.62 & -37.01 \\
Proposed (UMi) & -30.77 & -29.76 & -32.82 \\
Proposed (InH) & -29.45 & -28.77 & -31.78 \\
Proposed (CDL-A) & -29.38 & -28.57 & -31.94 \\
CsiNet-UPA & -15.01 & -14.85 & -15.54 \\ \bottomrule
\end{tabular}
}
\end{table}

All methods are evaluated on identical, held-out test sets.
The comparative performance is quantified using five primary metrics: mean NMSE in dB, 90th and 95th-percentile NMSE thresholds to assess worst-case reliability, CR, model footprint in MB, and average inference latency in milliseconds (ms) measured on standard UE hardware.

\subsection{Patch NMSE vs. CFR-NMSE}
To ensure a fair comparison, it is critical to clarify the distinction between the evaluation metrics used.
Let the full spatial-frequency CFR matrix be $\mathbf{H} \in \mathbb{C}^{N_\text{sc} \times N_\text{r} \times N_\text{c}}$.
CsiNet operates on delay-angular domain data, generated by applying a unitary 2D transform $\mathcal{F}(\cdot)$ such that the transformed channel is $\tilde{\mathbf{H}} = \mathcal{F}(\mathbf{H})$.
To achieve compression, CsiNet truncates this matrix, retaining only a dense sub-matrix patch $\tilde{\mathbf{H}}_\text{p}$ containing the dominant paths, while permanently discarding the out-of-patch high-delay energy $\tilde{\mathbf{H}}_\text{out}$.

Let the neural network's reconstruction error on the retained patch be $\mathbf{e}_\text{p} = \tilde{\mathbf{H}}_\text{p} - \hat{\tilde{\mathbf{H}}}_\text{p}$.
The reconstructed delay-angular matrix is zero-padded to restore the original dimensions, yielding $\hat{\tilde{\mathbf{H}}}_\text{padded} = [\hat{\tilde{\mathbf{H}}}_\text{p} ; \mathbf{0}]$.
The estimated CFR is then $\hat{\mathbf{H}} = \mathcal{F}^{-1}(\hat{\tilde{\mathbf{H}}}_\text{padded})$.
Because $\mathcal{F}$ is unitary, Parseval's theorem guarantees that the Frobenius norm of the reconstruction error is preserved across domains:
\begin{align}
\|\mathbf{H} - \hat{\mathbf{H}}\|_F^2 &= \|\tilde{\mathbf{H}} - \hat{\tilde{\mathbf{H}}}_\text{padded}\|_F^2 \\
&= \left\| \begin{bmatrix}
    \tilde{\mathbf{H}}_\text{p} - \hat{\tilde{\mathbf{H}}}_\text{p} \\
    \tilde{\mathbf{H}}_\text{out} - \mathbf{0}
\end{bmatrix} \right\|_F^2 \\
&= \|\mathbf{e}_\text{p}\|_F^2 + \|\tilde{\mathbf{H}}_\text{out}\|_F^2.
\end{align}
Most prior deep learning studies report the \textit{patch NMSE}, computed strictly over the truncated sub-matrix as $\|\mathbf{e}_\text{p}\|_F^2 / \|\tilde{\mathbf{H}}_\text{p}\|_F^2$.
In contrast, the true CFR-NMSE must account for the permanently discarded out-of-patch energy:
\begin{equation}
    \text{CFR-NMSE} = \frac{\|\mathbf{e}_\text{p}\|_F^2 + \|\tilde{\mathbf{H}}_\text{out}\|_F^2}{\|\tilde{\mathbf{H}}_\text{p}\|_F^2 + \|\tilde{\mathbf{H}}_\text{out}\|_F^2}.
\end{equation}
Consequently, as long as the neural network is successfully learning ($\|\mathbf{e}_\text{p}\|_F^2 < \|\tilde{\mathbf{H}}_\text{p}\|_F^2$) and truncation removes any non-zero energy ($\|\tilde{\mathbf{H}}_\text{out}\|_F^2 > 0$), the true CFR-NMSE is strictly larger (i.e., worse) than the patch NMSE.
In this work, all baselines are evaluated using the rigorous CFR-NMSE metric.

\subsection{CIR Structure Analysis}
Before evaluating reconstruction fidelity, we analyze the structural properties of the channel impulse responses across the datasets.
We visualized the CIR in environments with completely different propagation properties in Fig.~\ref{fig:cir_comparison}
As expected, the delays were longer in outdoor environments than in indoor environments.

\begin{figure}
    \centering
    \includegraphics[width=\if\numcol1 0.4 \else 0.85\fi\columnwidth]{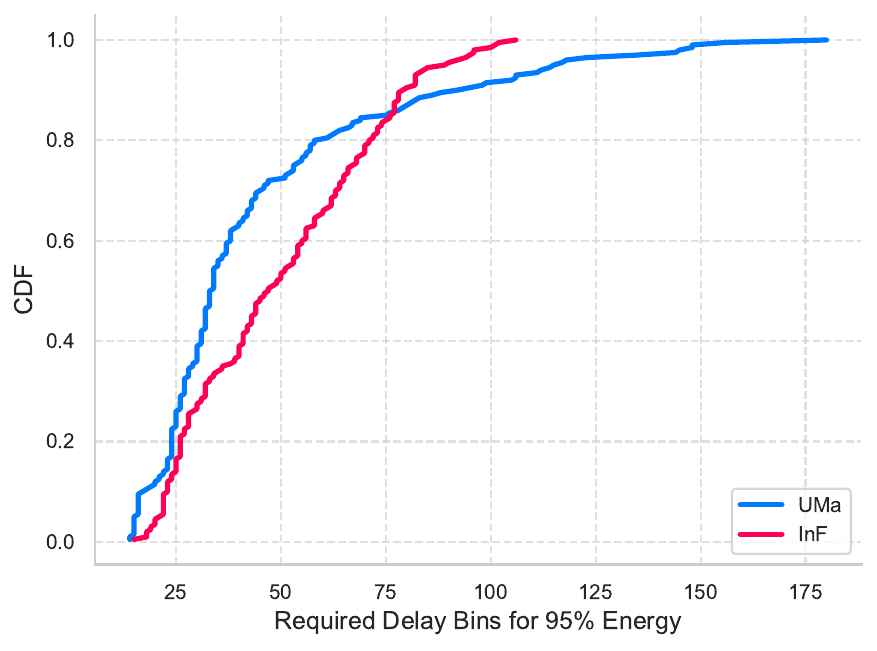}
    \caption{Cumulative distribution function (CDF) of the required number of delay bins to capture 95\% of the total channel energy in indoor and outdoor environments. The energy spread dictates the required scaling of the structural components.}
    \label{fig:energy_spread}
\end{figure}

Furthermore, the number of significant paths required for high-fidelity reconstruction differs fundamentally depending on the environment.
As demonstrated by the energy spread analysis in Fig.~\ref{fig:energy_spread}, indoor environments feature dense multipath scattering that spreads channel energy across numerous taps, necessitating a larger $K$ value to capture 95\% of the total channel energy~\cite{3gpp.38.901}.
In contrast, outdoor channels concentrate their energy into fewer, well-defined clusters, requiring a much smaller $K$.
Mathematically, the required number of paths $K$ is determined by the Power Delay Profile (PDP) energy threshold.
Letting $P(\tau)$ denote the expected power at delay $\tau$, the required number of discrete paths $K$ to capture a fraction $\eta$ (e.g., $\eta = 0.95$) of the total channel energy is implicitly defined by:
\begin{equation}
    \frac{\sum_{i=1}^K P(\tau_i)}{\sum_{\text{all } j} P(\tau_j)} \geq \eta,
\end{equation}
where the delays $\{\tau_i\}$ are ordered in descending order of power, such that $P(\tau_i) \geq P(\tau_{i+1})$ for all $i$.
Indoor environments typically exhibit a denser power delay profile $P(\tau)$ due to rich multipath scattering within physically bounded spaces, causing the median channel to require a substantially larger path count K to capture 95\% of the channel energy.
Conversely, while outdoor UMa environments are sparser on average, they lack hard physical boundaries and thus exhibit a heavy-tailed delay spread driven by distant unconstrained reflectors, demanding high $K$ strictly to capture worst-case channels.
Because the general base station deployment category (e.g., indoor versus outdoor) is known prior to operation, configuring an environment-appropriate maximum path count $K$ does not compromise the architecture's capacity to adapt to completely unseen environments in a zero-shot manner.

\if\numcol1
    \begin{figure}
        \centering
        \subfloat[]{\includegraphics[width=0.45 \linewidth]{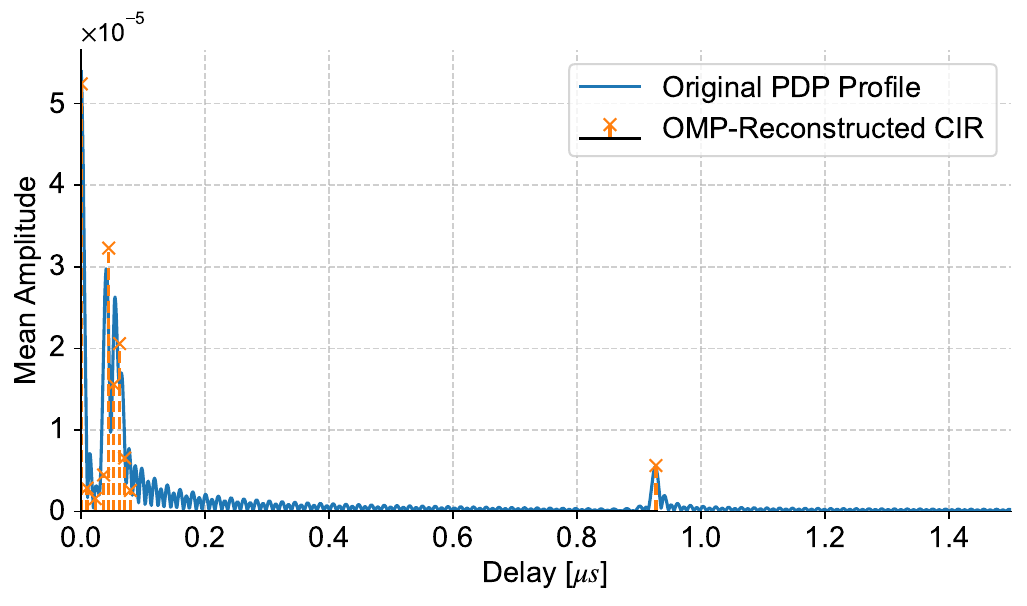}\label{subfig:outdoor_cir}}
        \subfloat[]{\includegraphics[width=0.45 \linewidth]{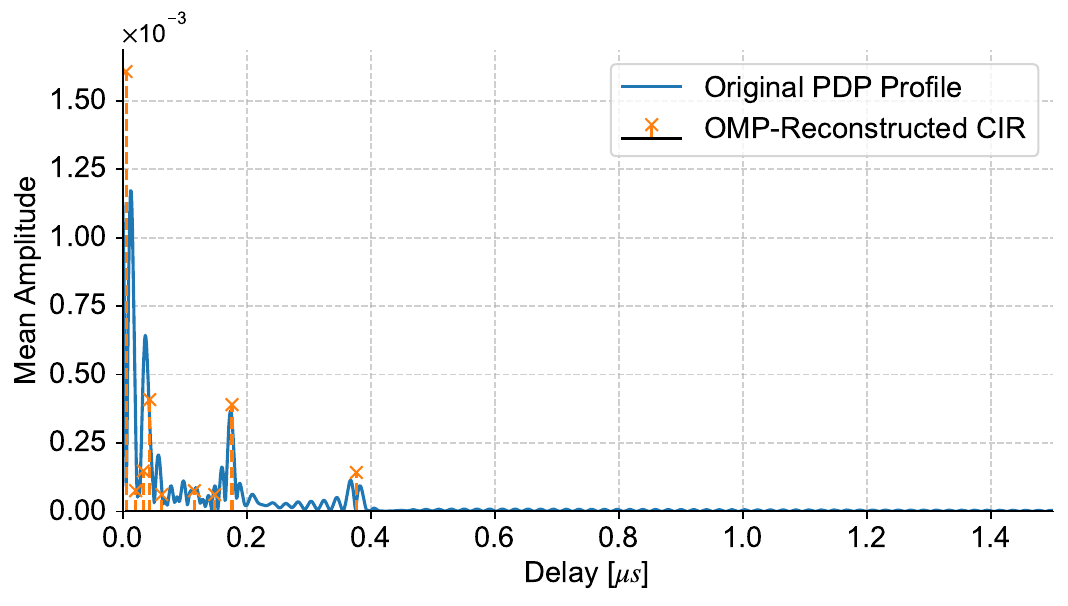}\label{subfig:indoor_cir}}
        \caption{CIR plots of different environments: (a) Outdoor environment; (b) Indoor environment.}
        \label{fig:cir_comparison}
    \end{figure}
\else
    \begin{figure}
        \centering
        \subfloat[]{\includegraphics[width=0.85 \linewidth]{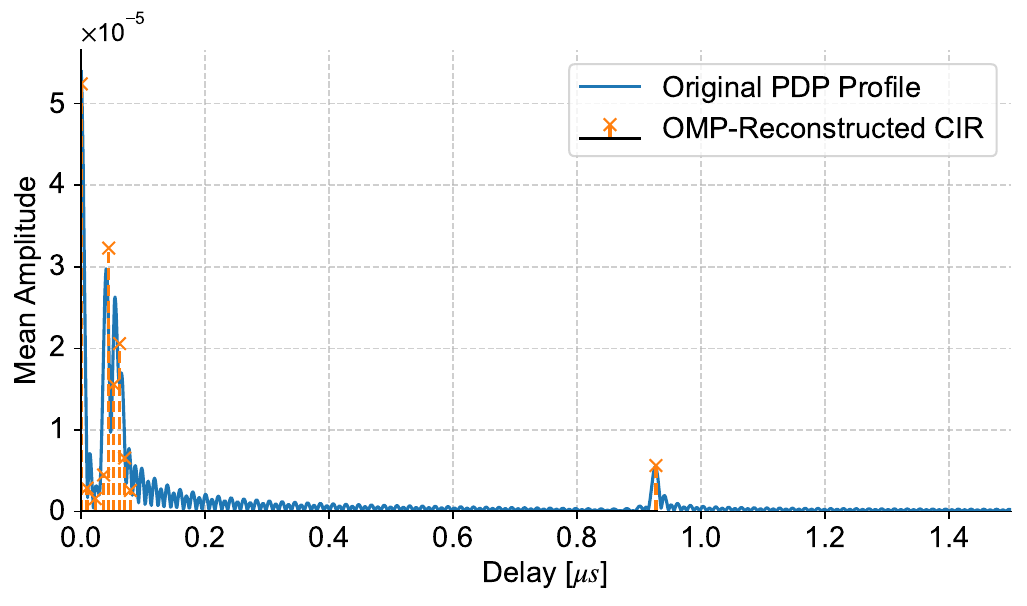}\label{subfig:outdoor_cir}}
        \hfil
        \subfloat[]{\includegraphics[width=0.85 \linewidth]{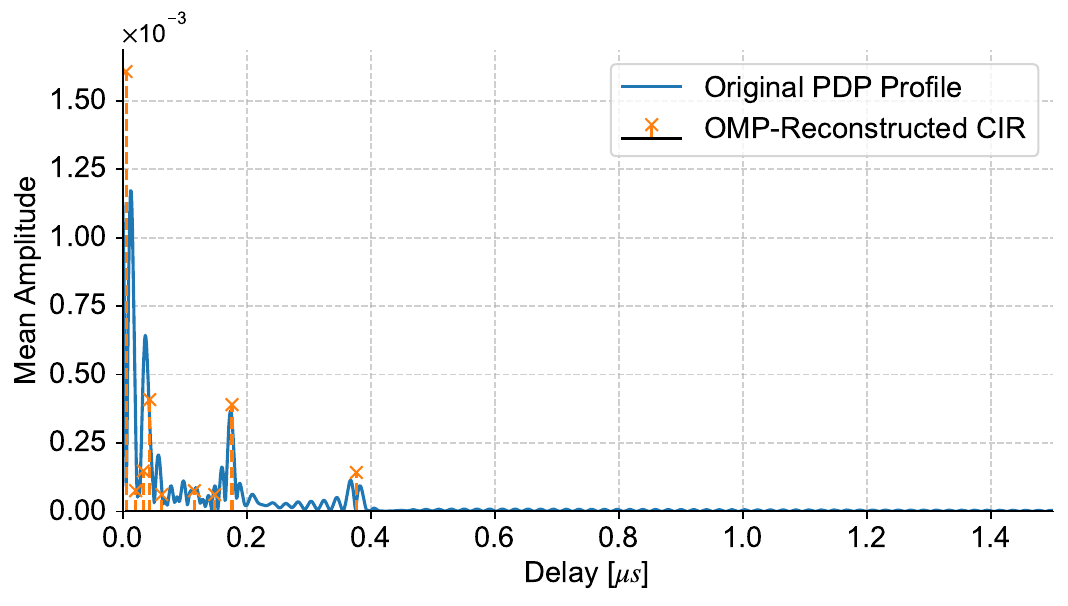}\label{subfig:indoor_cir}}
        \caption{Example CIR plots of different environments: (a) Outdoor environment; (b) Indoor environment.}
        \label{fig:cir_comparison}
    \end{figure}
\fi

\subsection{Reconstruction Quality}
\begin{figure}
    \centering
    \includegraphics[width=\if\numcol1 0.5 \else 0.85\fi\columnwidth]{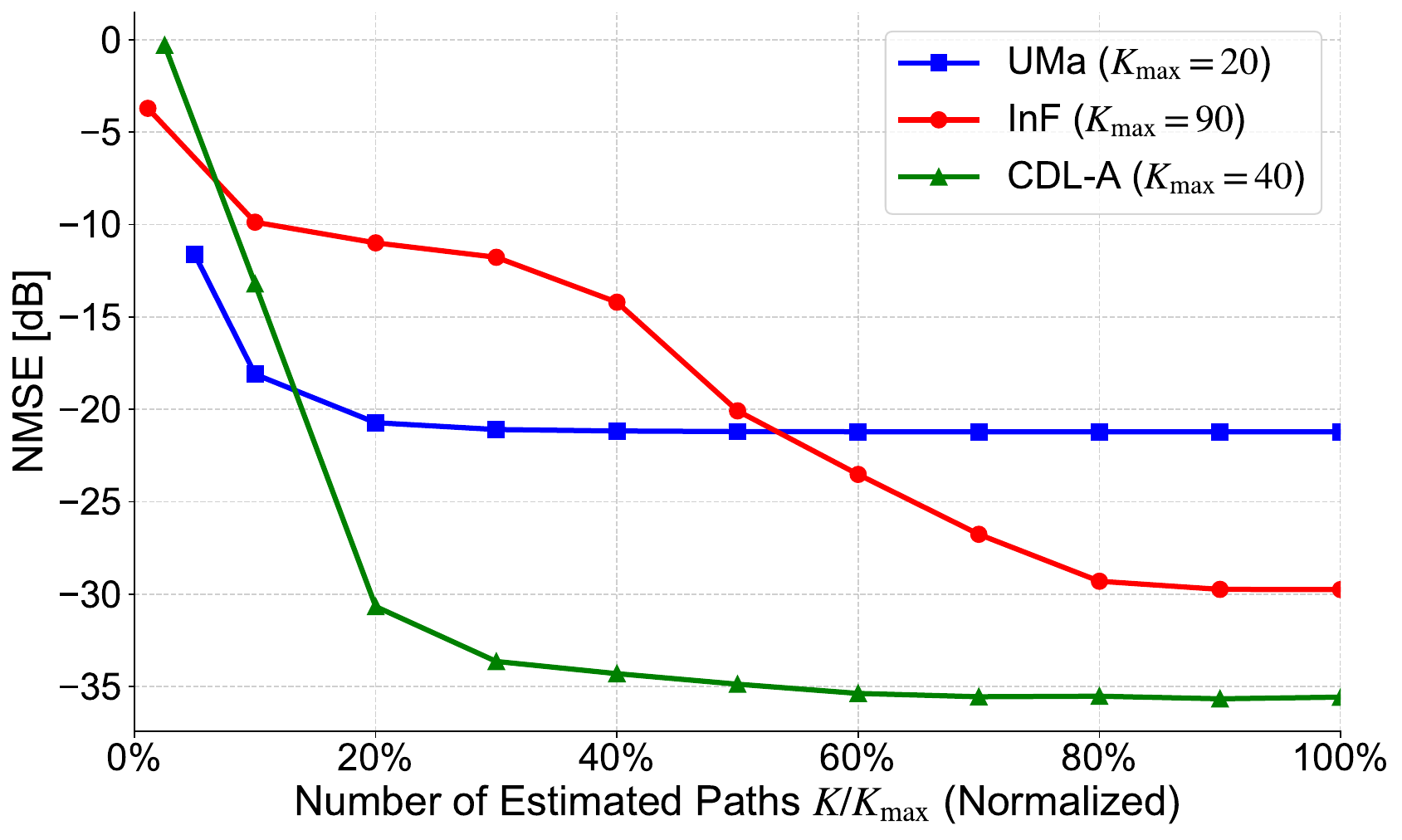}
    \caption{NMSE distributions of the proposed method under different environments.}
    \label{fig:nmse_vs_paths}
\end{figure}

Fig.~\ref{fig:nmse_vs_paths} illustrates the mean CFR-NMSE as a function of the number of estimated paths $K$.
Across all environments, the reconstruction quality improves monotonically as $K$ increases, eventually saturating at a point corresponding to the physical path count of the environment.
At the optimal $K$ for each dataset, the proposed TAP method achieves mean CFR-NMSE values ranging from $-21.2$~dB in UMa and $-29.8$~dB in InF to $-35.7$~dB in CDL-A.

\begin{figure}
    \centering
    \subfloat[]{\includegraphics[width=\if\numcol1 0.3 \else 0.75\fi\columnwidth]{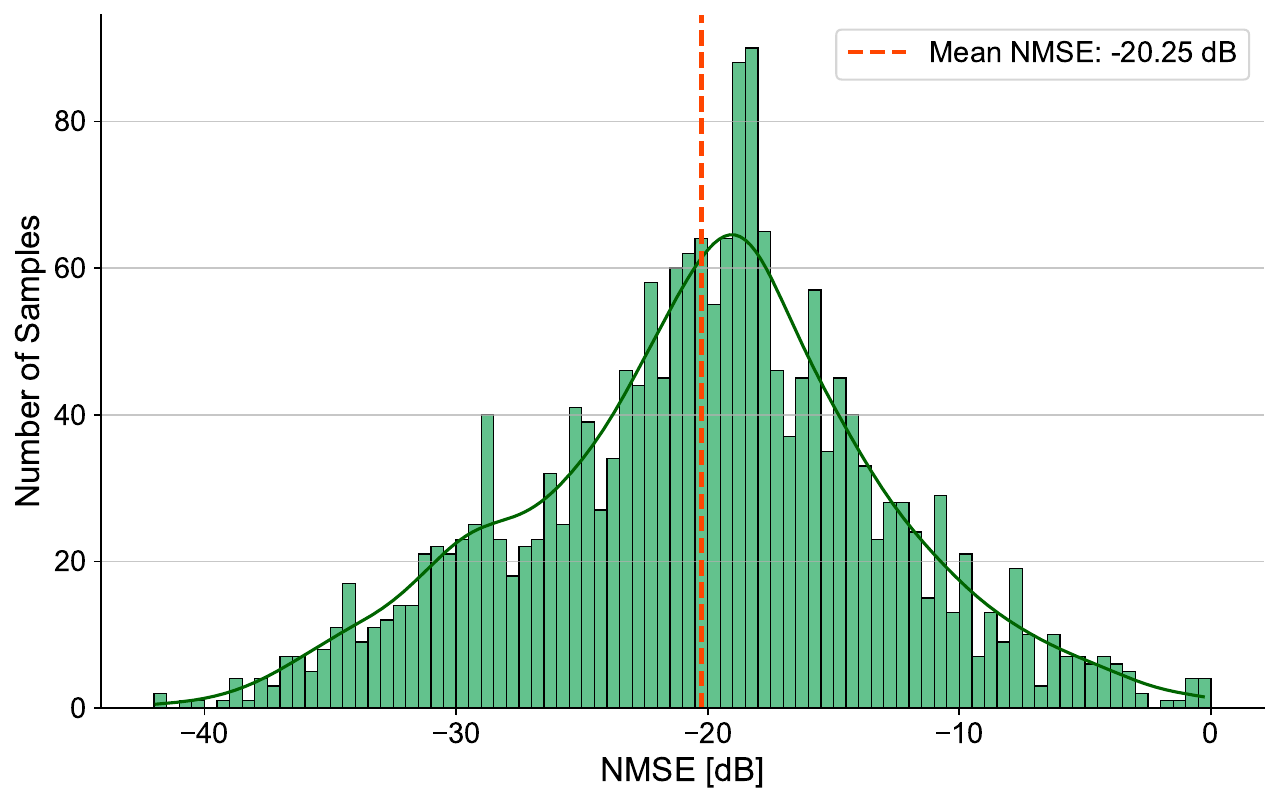}\label{subfig:nmse_dist_outdoor}}
    \hfil
    \subfloat[]{\includegraphics[width=\if\numcol1 0.3 \else 0.75\fi\columnwidth]{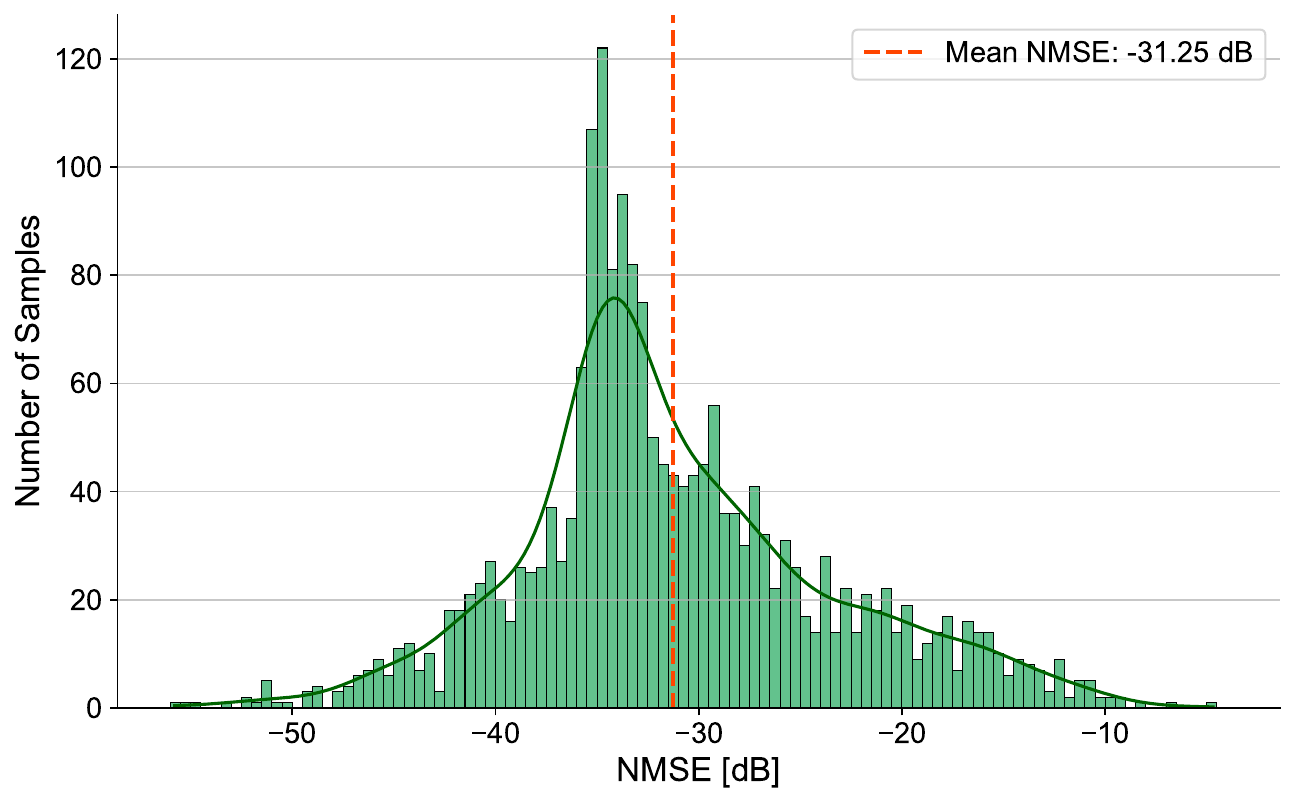}\label{subfig:nmse_dist_indoor}}
    \hfil
    \subfloat[]{\includegraphics[width=\if\numcol1 0.3 \else 0.75\fi\columnwidth]{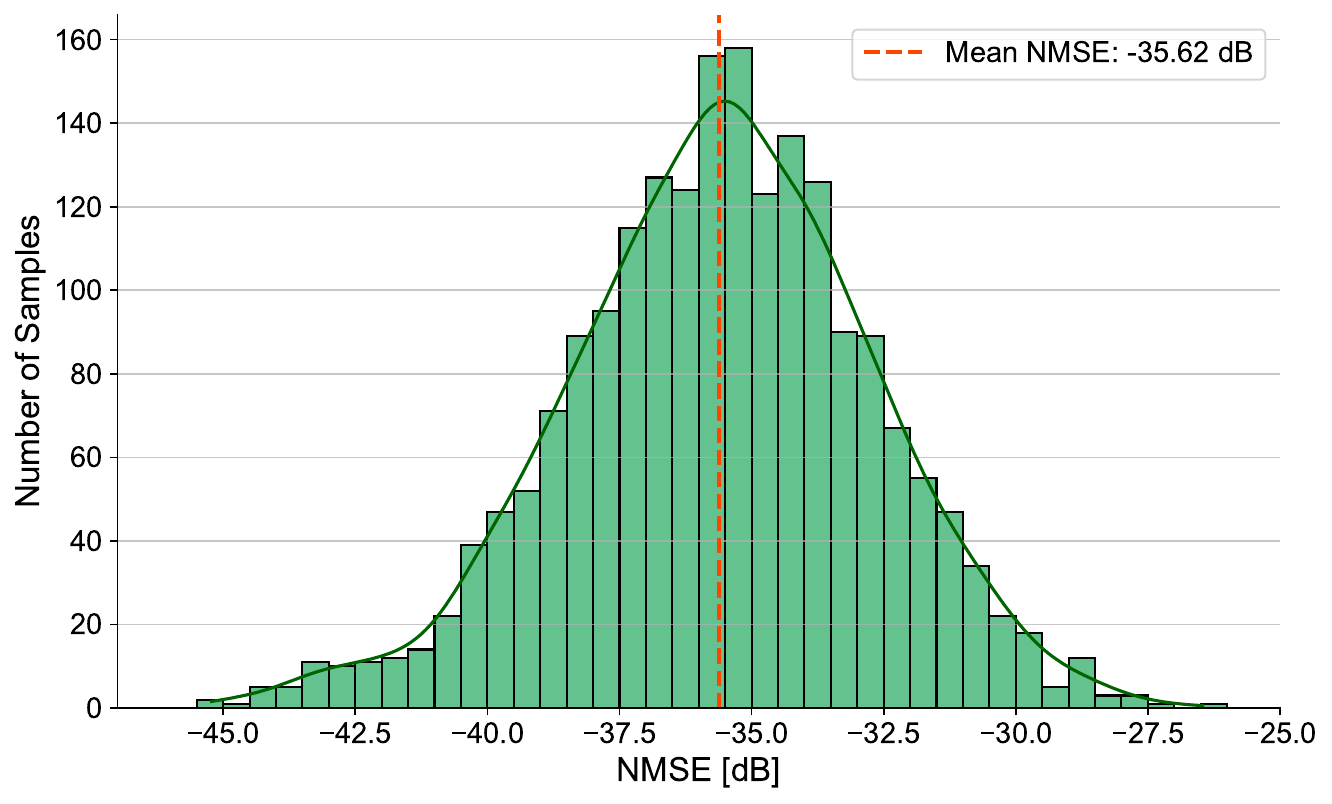}\label{subfig:nmse_dist_sionna}}
    \caption{NMSE distributions of the proposed method under different environments; (a) Outdoor environment; (b) Indoor environment; (c) CDL-A environment.}
    \label{fig:nmse_distribution}
\end{figure}

Beyond average performance, statistical robustness is critical for reliable communication links.
The CFR-NMSE distribution histograms in Fig.~\ref{fig:nmse_distribution} and the percentile data in Table~\ref{tab:temp_tab5} reveal highly concentrated error distributions.
For instance, over 92\% of samples in the UMa dataset achieve better than $-10$~dB CFR-NMSE, while the InF and CDL-A datasets show over 99.6\% of samples surpassing this threshold.
This confirms that TAP provides reliable per-sample reconstruction rather than just favorable average metrics.

\begin{table}
\centering
\caption{Performance Comparison of the Proposed Method in Different Environments}
\label{tab:temp_tab5}
\adjustbox{width=\if\numcol1 0.7 \else 1 \fi\linewidth}{
\begin{tabular}{@{}cccccc@{}}
\toprule
\multirow{2}{*}{\textbf{Environment}} & \multirow{2}{*}{\textbf{CR}} & \multirow{2}{*}{\textbf{-10~dB Prob. [\%]}} & \multicolumn{3}{c}{\textbf{CFR NMSE {[}dB{]}}} \\ \cmidrule(l){4-6} 
 &  &  & $P_{90}$ & $P_{95}$ & Mean \\ \midrule
UMa & 0.0124 & 92.71 & -11.38 & -8.29 & -20.25 \\
InF & 0.0558 & 99.65 & -20.31 & -16.38 & -31.25 \\
UMi & 0.0248 & 100.0 & -27.07 & -26.95 & -50.71 \\
InH & 0.0248 & 100.0 & -36.11 & -36.09 & -38.62 \\
CDL-A & 0.0248 & 100.0 & -32.15 & -31.15 & -35.76 \\ \midrule
\multicolumn{6}{l}{* $P_{90}$ and $P_{95}$ denote the 90th and 95th percentile NMSE thresholds, respectively.}
\end{tabular}
}
\end{table}

\subsection{Compression-Quality Pareto Efficiency}

\begin{figure}
    \centering
    \subfloat[]{\includegraphics[width=\if\numcol1 0.4 \else 0.85\fi\columnwidth]{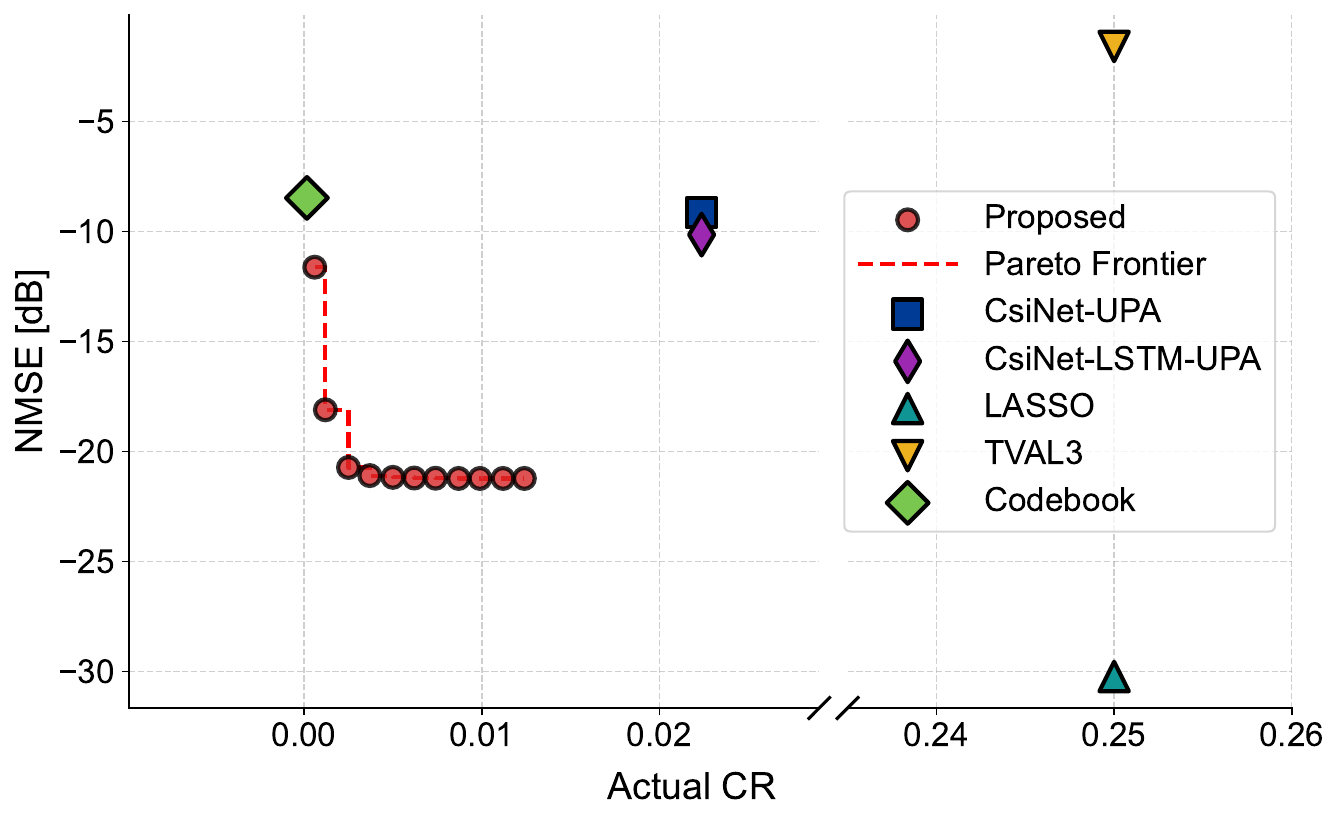}\label{subfig:pareto_outdoor}}
    \hfil
    \subfloat[]{\includegraphics[width=\if\numcol1 0.3 \else 0.85\fi\columnwidth]{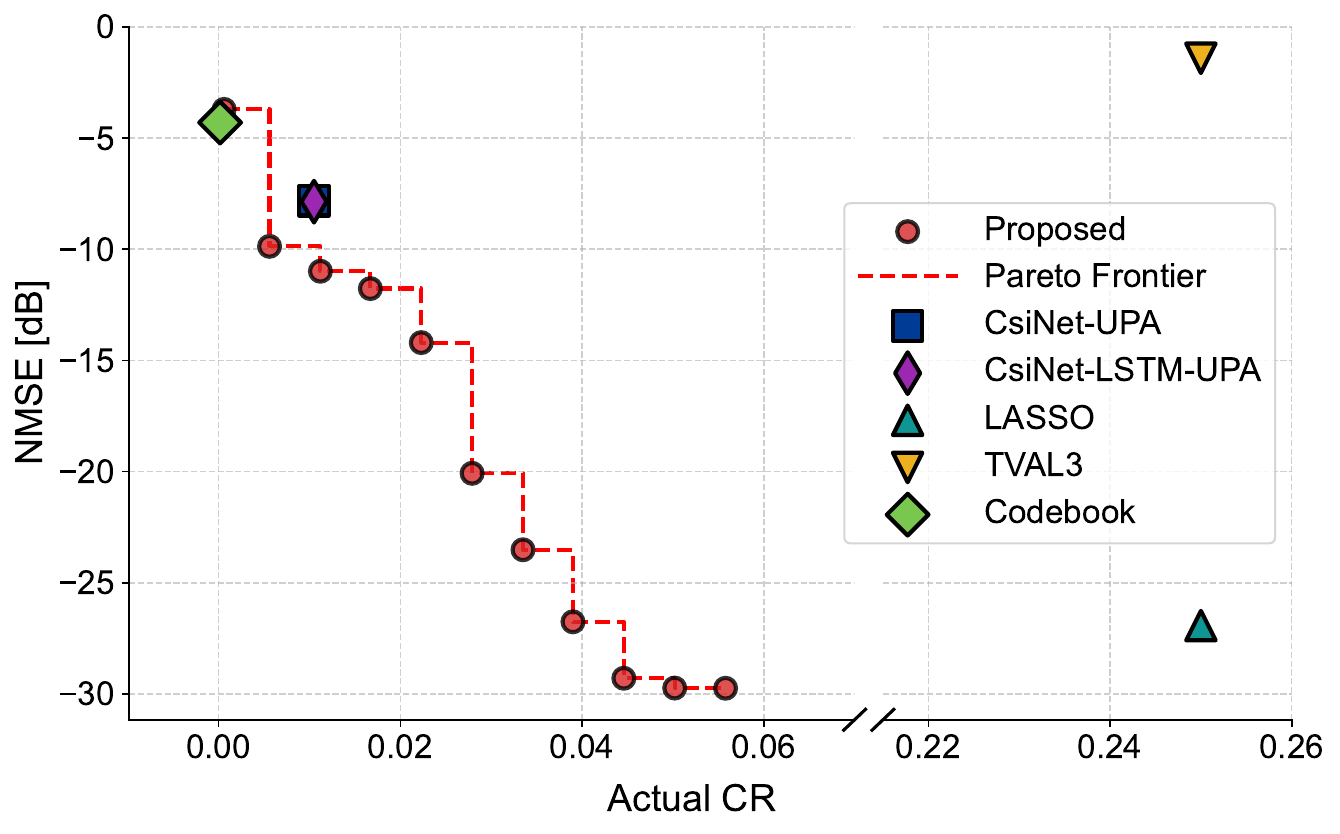}\label{subfig:pareto_indoor}}
    \caption{Pareto efficiency graph of the proposed method under different environments; (a) UMa environment; (b) InF environment.}
    \label{fig:pareto}
\end{figure}

The trade-off between CR and reconstruction fidelity is the key metric for CSI feedback.
The Pareto frontier analysis presented in Fig.~\ref{fig:pareto} maps this trade-off for both UMa and InF environments.
The proposed TAP framework strictly dominates the baseline methods, defining a new Pareto frontier.
For example, in the outdoor UMa environment, the proposed model achieves an NMSE of -21.22~dB at a CR of 0.0124. Even when using a significantly smaller CR of 0.0012, the proposed model achieves an NMSE of -18.10~dB, which is still substantially lower than the CsiNet-UPA's NMSE of -9.14~dB at a much larger CR of 0.0224.
Furthermore, in the InF environment, although CsiNet-UPA requires a CR of 0.0105 to achieve an NMSE of -7.82~dB, the proposed method achieves a lower NMSE of -10.99~dB at a similar CR of 0.0112, and can further push the NMSE down to -20.08~dB at a CR of 0.0279.
Overall, TAP consistently outperforms DL-based baselines across all evaluated scenarios, as summarized in Table~\ref{tab:complexity}.

\subsection{Cross-Domain Generalization}
\begin{figure}
    \centering
    \includegraphics[width=\if\numcol1 0.3 \else 0.7\fi\columnwidth]{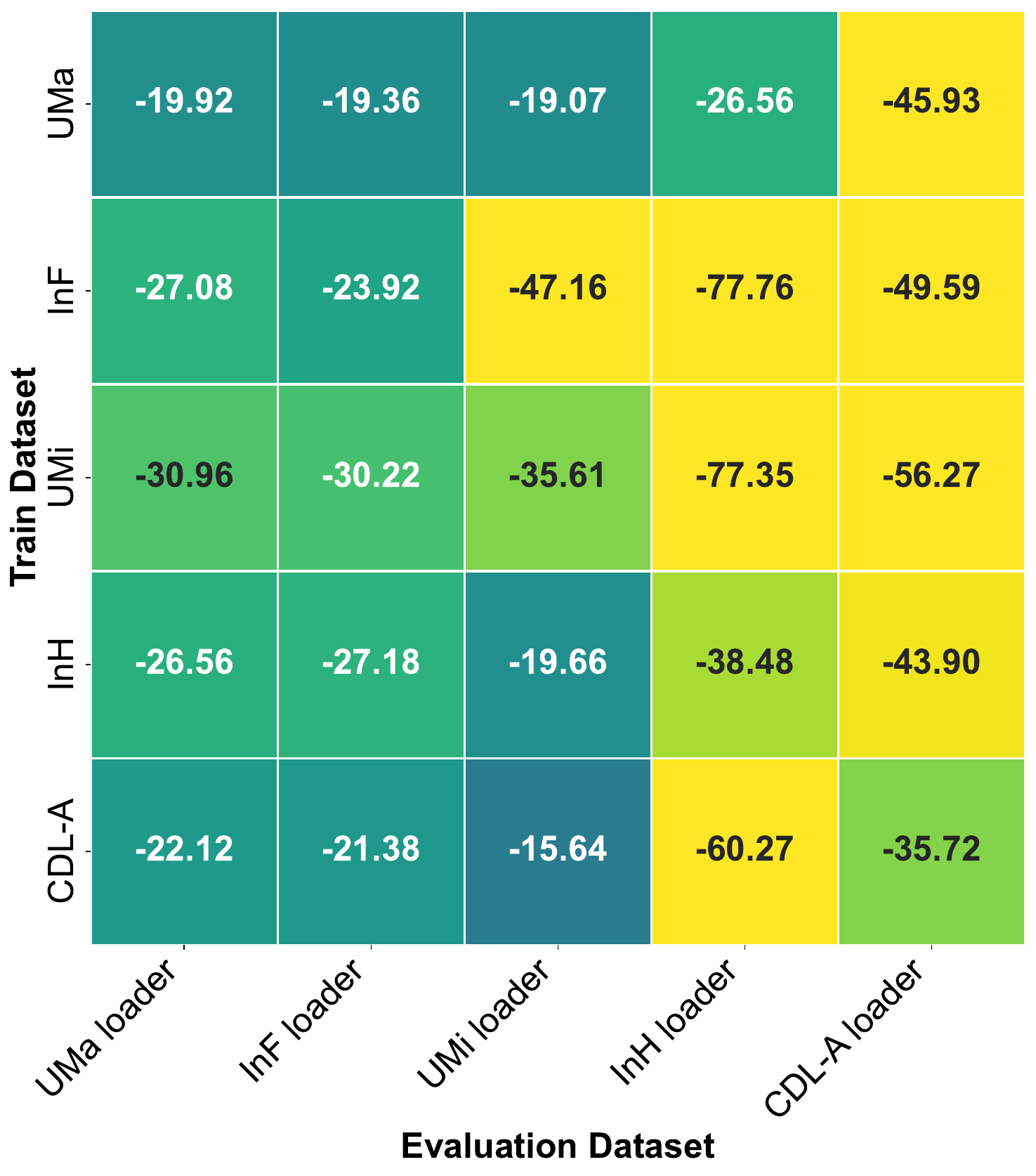}
    \caption{Cross-domain zero-shot generalization performance matrix between different environments.}
    \label{fig:cross_data_matrix}
\end{figure}

A major vulnerability of existing neural compressors is their inability to adapt to unseen environments.
We evaluate zero-shot cross-environment transfer by training TAP on a single source dataset and evaluating it on the remaining four target datasets without fine-tuning.
The resulting $5 \times 5$ CFR-NMSE matrix in Fig.~\ref{fig:cross_data_matrix} demonstrates strong generalization.
For example, a model trained on the InF environment achieves $-47.16$~dB CFR-NMSE on the outdoor UMi dataset, while a UMa-trained model yields $-26.56$~dB on the InH dataset.
This zero-shot transfer is enabled by TAP's reliance on physical spatial-delay structure and inference-time time-of-flight alignment, which decouples the neural weights from environment-specific statistics.

\subsection{Antenna Geometry Scalability}
\begin{figure*}
    \centering
    \subfloat[]{\includegraphics[width=0.3\linewidth]{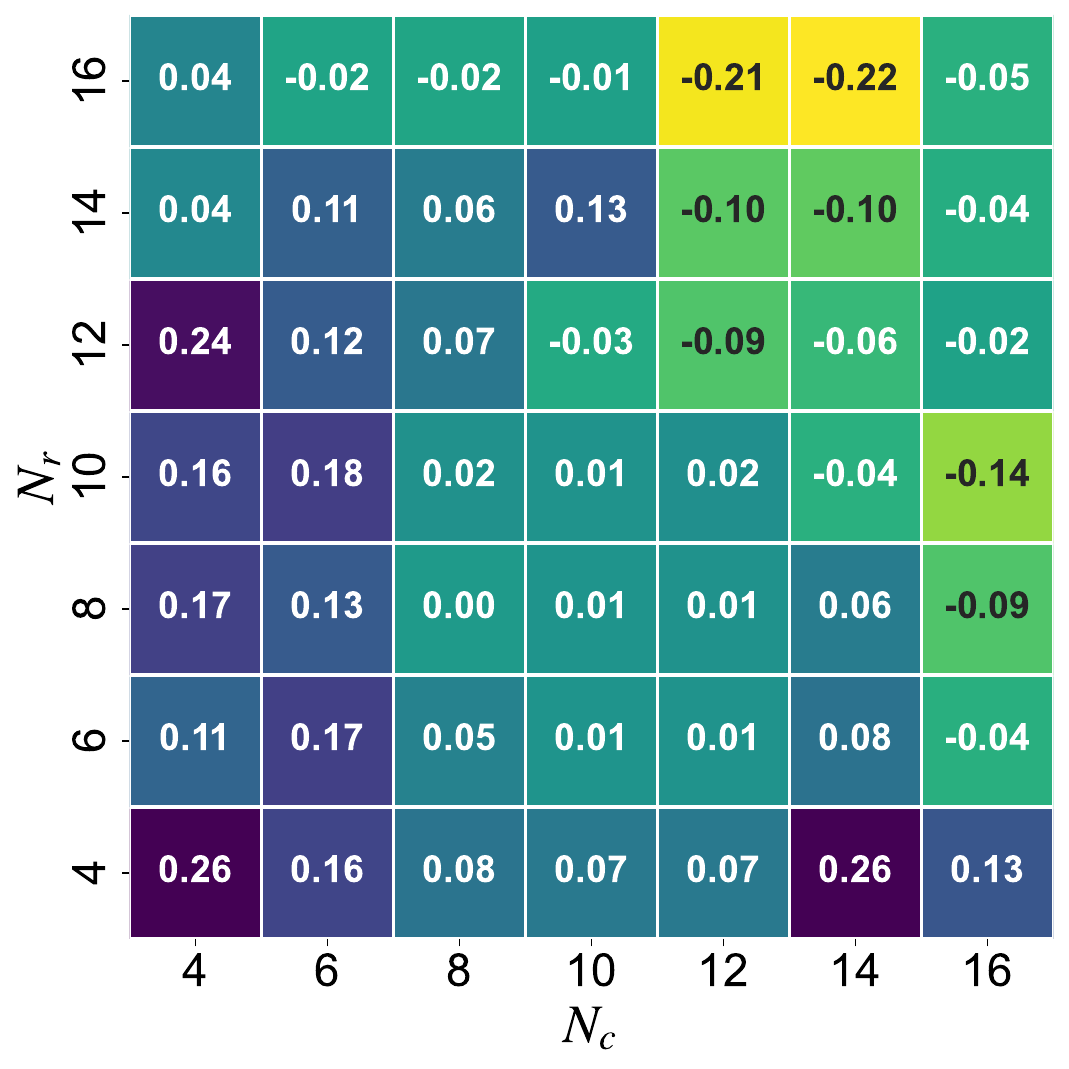}\label{subfig:scalability_outdoor}}
    \hfil
    \subfloat[]{\includegraphics[width=0.3\linewidth]{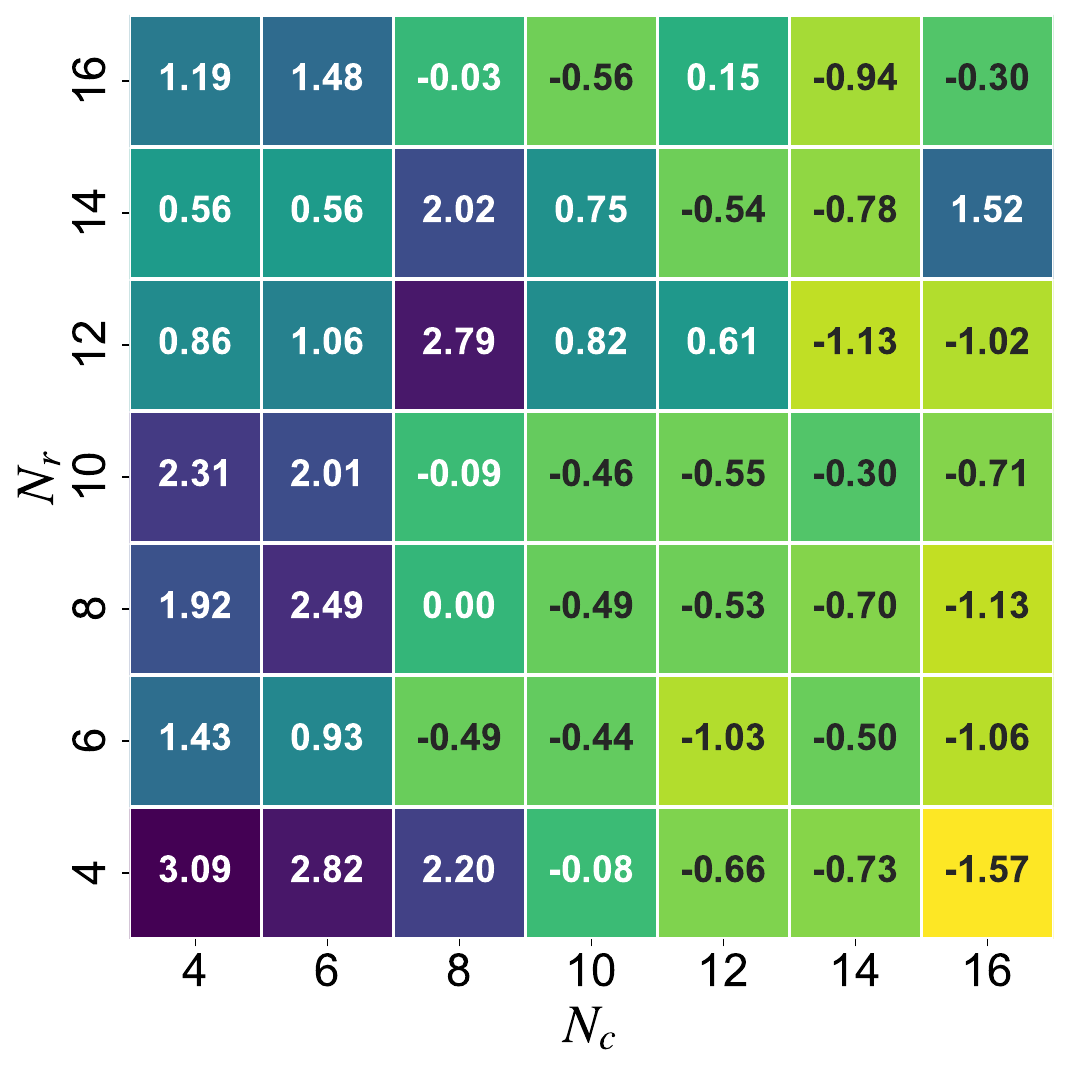}\label{subfig:scalability_indoor}}
    \hfil
    \subfloat[]{\includegraphics[width=0.3\linewidth]{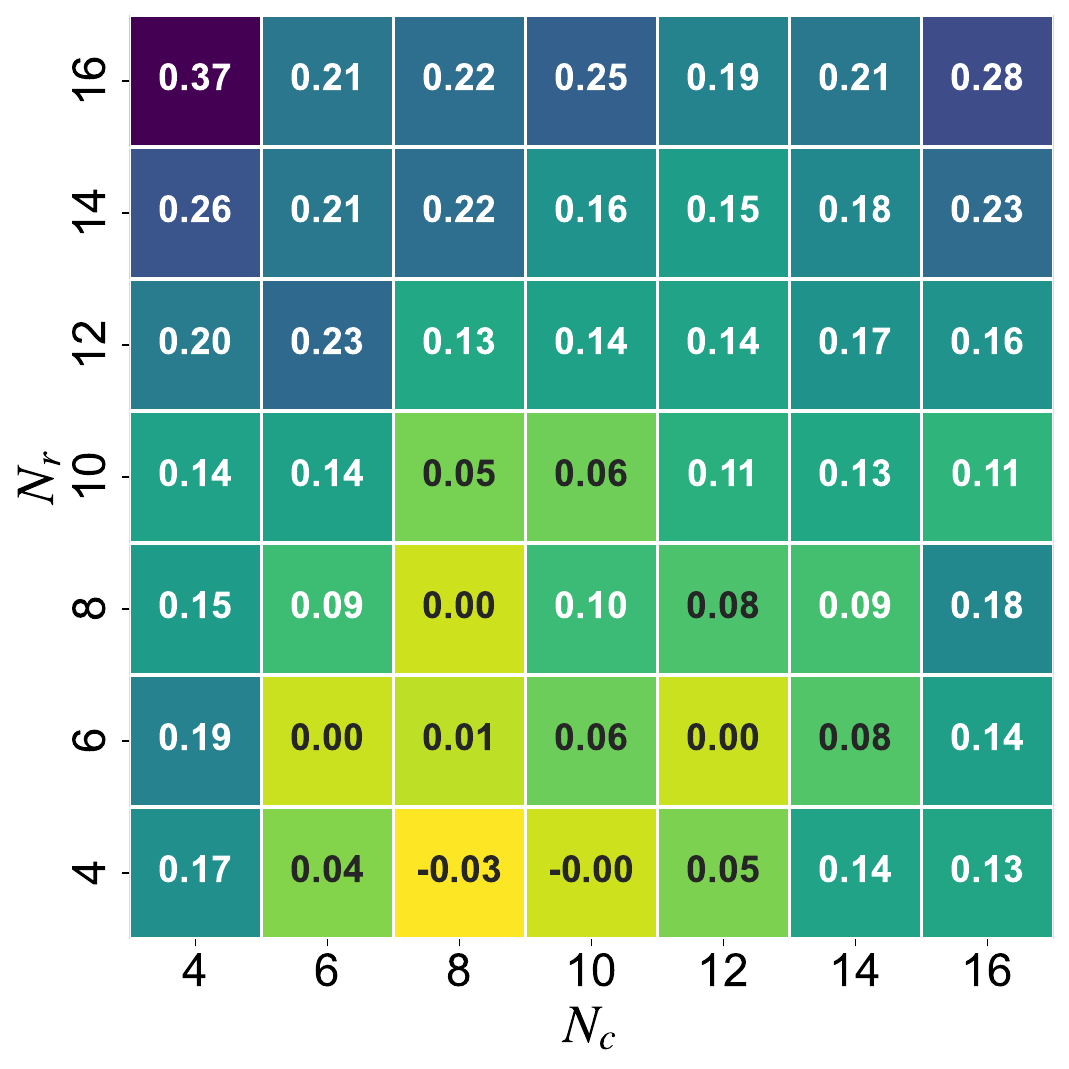}\label{subfig:scalability_sionna}}
    \caption{Geometric scalability evaluation: NMSE performance across heterogeneous antenna array geometries; (a) Outdoor-trained model; (b) Indoor-trained model; (c) Sionna-PHY-trained model.}
    \label{fig:cross_antenna_matrix}
\end{figure*}

TAP also exhibits zero-shot cross-antenna generalization.
The depthwise separable convolutions and path-agnostic batch folding make the architecture independent of the BS antenna array dimensions.
As shown in Fig.~\ref{fig:cross_antenna_matrix}, a model trained on a specific UPA geometry can be deployed on unseen configurations (e.g., $4 \times 16$, $16 \times 4$, or $1 \times 32$ arrays).
Across all three primary training environments, the CFR-NMSE degradation when transferring to an unmatched geometry is less than 3~dB compared to the matched-geometry baseline, confirming spatial scalability.

We also note that Table~\ref{tab:ula_comparison} also shows the generalization performance of TAP.
All TAP models were trained in different environment and antenna configurations, but they all outperformed the CsiNet-UPA, which was trained using the ULA dataset.

\subsection{Complexity Analysis}

\begin{table}[t]
\centering
    \caption{Computation and NMSE Performance Comparison of Each Method in Different Environments (Default: UMa)}
    \label{tab:complexity}
    \adjustbox{width=\if\numcol1 0.7 \else 1 \fi\linewidth}{
    \begin{tabular}{@{}cccccc@{}}
    \toprule
    \textbf{Method (Environment)} & \textbf{CR} & \textbf{\begin{tabular}[c]{@{}c@{}}Model size\\ {[}MB{]} \end{tabular}} & \textbf{\begin{tabular}[c]{@{}c@{}}Avg. Inf. Time\\ {[}ms{]}\end{tabular}} & \textbf{Avg. GFLOPs} & \textbf{\begin{tabular}[c]{@{}c@{}}Mean NMSE\\ {[}dB{]}\end{tabular}} \\ \midrule
    Proposed (UMa) & 0.0124 & 0.9921 & 0.6239 & 0.1527 & -21.22 \\
    Proposed (InF) & 0.0558 & 0.4504 & 0.4824 & 0.0325 & -29.75 \\
    Proposed (CDL-A) & 0.0248 & 0.9921 & 0.6433 & 0.0762 & -35.57 \\
    Proposed (UMi) & 0.0248 & 0.9921 & 0.6228 & 0.1527 & -35.76 \\
    Proposed (InH) & 0.0248 & 0.4504 & 0.4774 & 0.0325 & -38.42 \\
    CsiNet-UPA (UMa) & 0.0224 & 657.1579 & 0.7625 & 0.2634 & -8.74 \\
    CsiNet-UPA (InF) & 0.0105 & 144.5797 & 0.3614 & 0.0807 & -2.99 \\
    CsiNet-UPA (CDL-A) & 0.0071 & 66.1912 & 0.3656 & 0.0463 & -15.39 \\
    OMP ($K=10$) & 0.0024 & N/A & 630.15 & N/A & -31.16 \\
    LASSO (CR=0.25) & 0.2500 & N/A & 88648.56 & N/A & -30.22 \\
    TVAL3 (CR=0.25) & 0.2500 & N/A & 1301.37 & N/A & -1.59 \\
    3GPP Type-1 Codebook & 0.00008 & N/A & 14.81 & N/A & -5.95 \\
    3GPP Type-2 Codebook & 0.00016 & N/A & 23.95 & N/A & -8.47 \\ \bottomrule
    \multicolumn{6}{l}{* The inference time and the FLOPs include pre- and post-processing.}
    \end{tabular}
    }
\end{table}

Real-time FDD CSI feedback mandates strict operational efficiency at the UE.
As documented in Table~\ref{tab:complexity}, TAP achieves sub-millisecond inference latencies ($0.45$ to $0.60$~ms) while consuming less than $1$~MB of memory ($0.45$ to $0.99$~MB) and operating at $0.03$ to $0.15$~GFLOPs.
This represents a 2700 times acceleration over iterative OMP and a 6400 times acceleration over LASSO.

Crucially, TAP provides massive memory savings over deep learning baselines.
While the original CsiNet was designed with a relatively small memory footprint for ULAs with limited subcarriers, extending its architecture to support UPAs and realistic wideband grids drastically inflates its size.
Because the modified CsiNet-UPA relies on dense fully connected layers that scale quadratically with the spatial dimensions and retained delay patch sizes, its memory footprint reaches $657.2$~MB for the outdoor UMa environment and $144.6$~MB for the indoor InF environment.
In contrast, TAP's path-agnostic architecture keeps the model footprint under $1$~MB across all scenarios.
This represents up to a 660 times reduction in model size, making TAP uniquely suited for memory-constrained UE deployments without sacrificing CFR-NMSE performance.

The speed of TAP stems from offloading the iterative search to a lightweight CNN, leaving only a single closed-form LS projection.
The computational complexity of computing the complex amplitudes $\mathbf{\hat{X}} = (\mathbf{A}^H \mathbf{A} + \lambda \mathbf{I})^{-1} \mathbf{A}^H \mathbf{H}_\text{tgt}$ is dominated by matrix multiplications and the inversion of a small $K \times K$ matrix.
For complex-valued matrices, constructing $\mathbf{A}^H \mathbf{A}$ requires $8 K^2 N_\text{sc}$ FLOPs, projecting the target $\mathbf{A}^H \mathbf{H}_\text{tgt}$ requires $8 K N_\text{sc} N_\text{ant}$ FLOPs, the complex matrix inversion takes approximately $\frac{16}{3} K^3$ FLOPs, and the final multiplication takes $8 K^2 N_\text{ant}$ FLOPs.
The total FLOP count for the algebraic recovery is $\text{FLOPs}_{\text{LS}} = 8 \left( K^2 N_\text{sc} + K N_\text{sc} N_\text{ant} + \frac{2}{3} K^3 + K^2 N_\text{ant} \right)$.
For typical parameters (e.g., $K=20$, $N_\text{sc}=1620$, $N_\text{ant}=64$), this evaluates to approximately $18$~MFLOPs.
Including the DelayNet passes, the entire pipeline remains under $153$~MFLOPs.

\subsection{System-Level Spectral Efficiency}
\begin{figure}
    \centering
    \includegraphics[width=\if\numcol1 0.5 \else 0.8\fi\columnwidth]{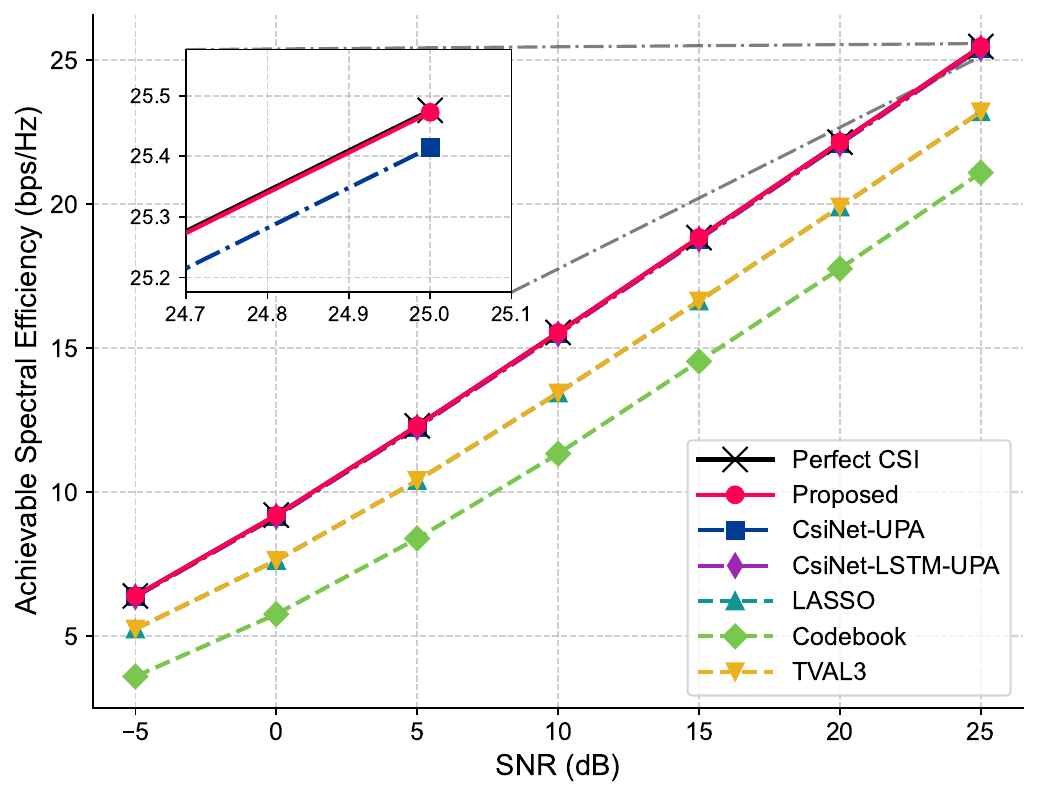}
    \caption{Downlink achievable spectral efficiency (bps/Hz) in the CDL-A environment using a linear precoder derived from the reconstructed CSI.}
    \label{fig:spectral_efficiency}
\end{figure}

To evaluate the downstream impact of the proposed CSI compression framework, we map the reconstructed CSI to achievable spectral efficiency.
Using the reconstructed CFRs to synthesize a linear spatial precoder at the base station, Fig.~\ref{fig:spectral_efficiency} plots the resulting single-user MIMO capacity.
Compared to the theoretical upper bound of perfect CSI, TAP exhibits only marginal capacity degradation across the entire SNR regime, tightly tracking the optimal curve.
In contrast, baselines such as CsiNet-UPA introduce capacity ceilings due to amplitude and phase distortion, translating their poor CFR-NMSE performance into throughput losses.
This confirms that TAP provides reliable physical channel representations capable of supporting high-order modulation schemes.

\subsection{Robustness to Payload Quantization}
\begin{figure}
    \centering
    \includegraphics[width=\if\numcol1 0.5 \else 0.8\fi\columnwidth]{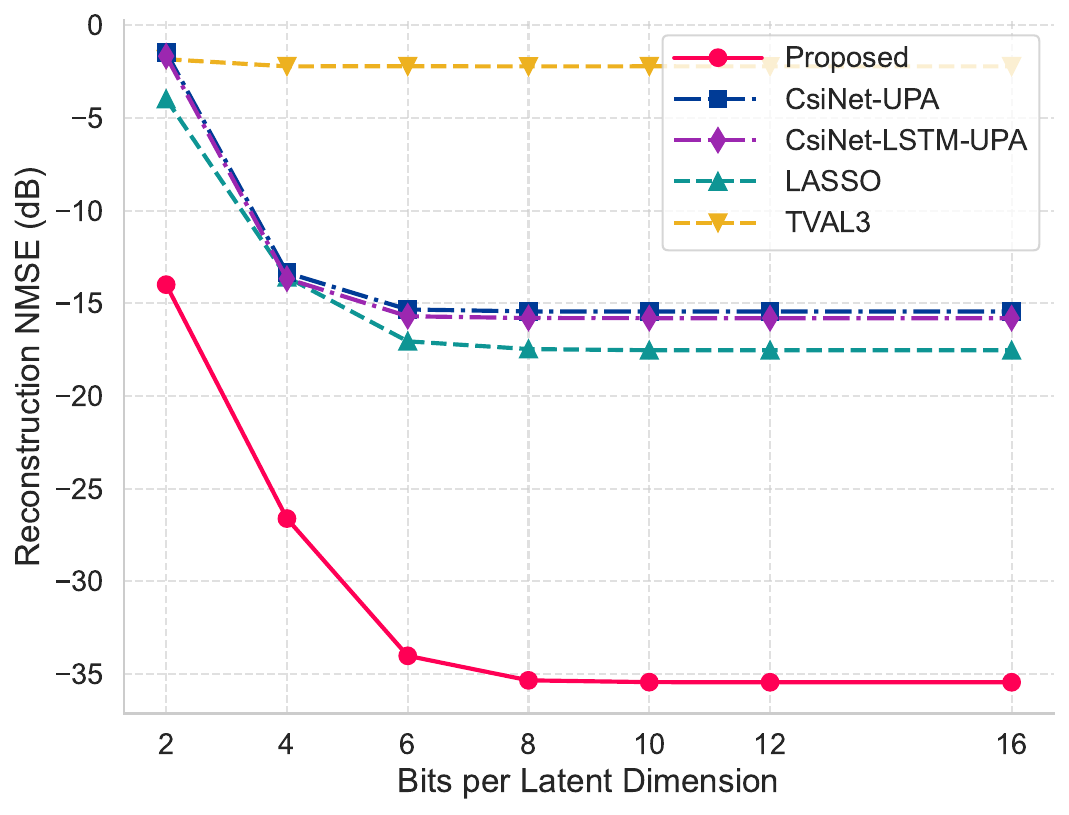}
    \caption{Impact of payload quantization on reconstruction NMSE across different bit-widths for the CDL-A dataset. TAP demonstrates high robustness to harsh variable bit-width quantization compared to traditional baseline autoencoders.}
    \label{fig:quantization_impact}
\end{figure}

In practical communication systems, the continuous representations must be quantized into a finite bitstream before transmission.
To evaluate the robustness of TAP against quantization noise, we apply a uniform scalar quantizer to the extracted spatial amplitudes across variable bit-widths.
As illustrated in Fig.~\ref{fig:quantization_impact}, TAP exhibits a steeper degradation slope compared to the CsiNet-UPA baseline as the bit-width decreases.
However, this steepness is a mathematical consequence of TAP's exceptionally low initial error floor.
Because the total reconstruction error is bounded by the sum of the model's inherent prediction error and the quantization noise, the baseline's inherently high model error ($-17.5$~dB NMSE of LASSO and $-15.4$~dB NMSE of CsiNet-UPA) masks the quantization noise at higher bit-widths, making its curve appear artificially robust.
In contrast, TAP's high-fidelity unquantized reconstruction (approximately $-35.4$~dB) exposes the standard 6~dB-per-bit quantization degradation linearly.
Despite this steeper slope, TAP strictly upper-bounds the baseline's performance, maintaining lower absolute CFR-NMSE even at resolutions as low as 4 to 6 bits per dimension.

\subsection{Robustness to CSI-RS Estimation Noise}
\begin{figure*}
    \centering
    \subfloat[]{\includegraphics[width=0.3\linewidth]{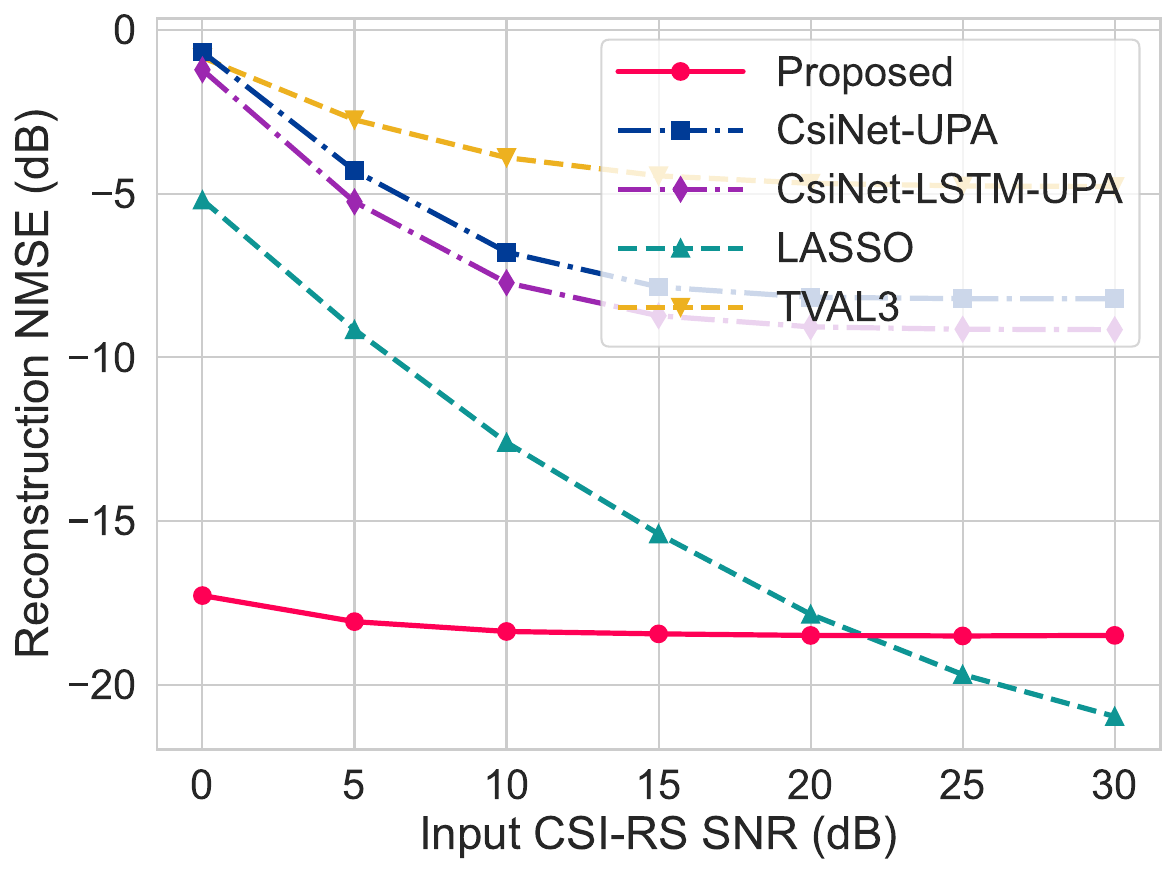}\label{subfig:noise_outdoor}}\hfil
    \subfloat[]{\includegraphics[width=0.3\linewidth]{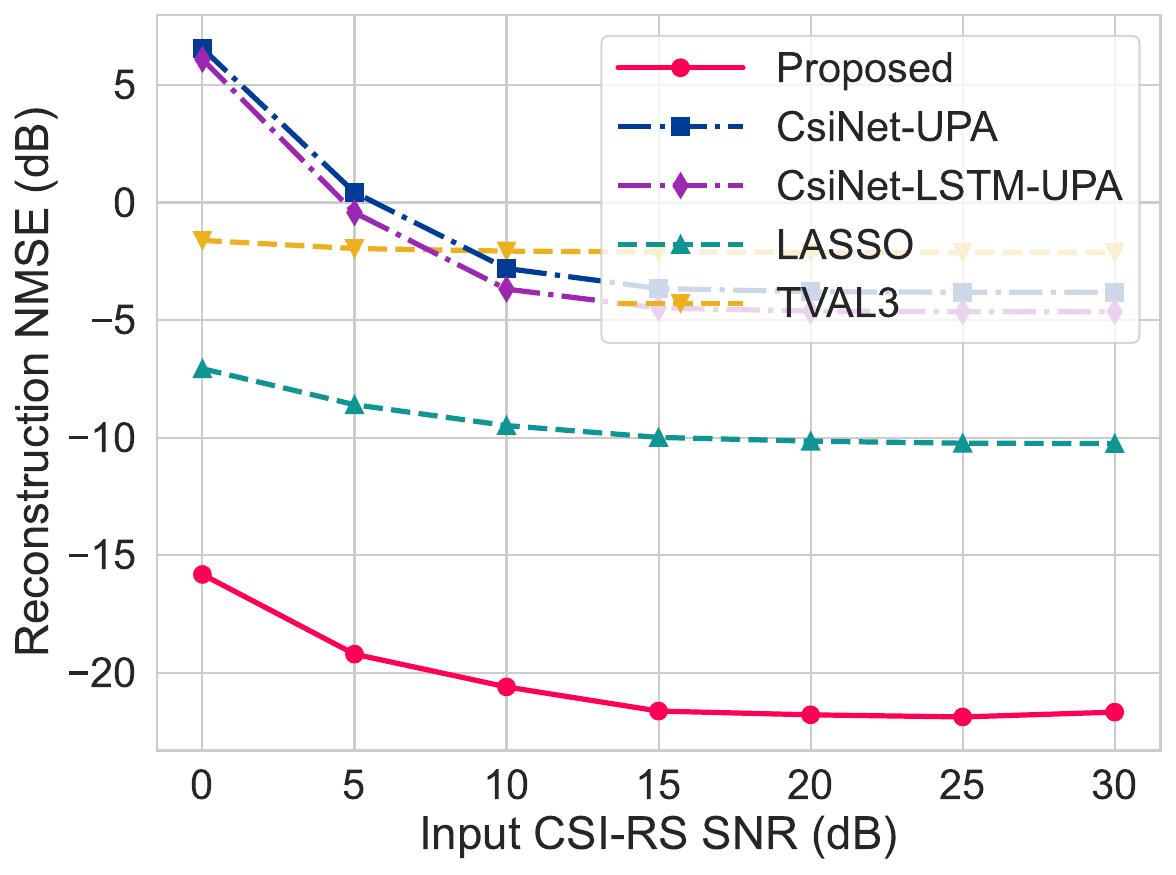}\label{subfig:noise_indoor}}\hfil
    \subfloat[]{\includegraphics[width=0.3\linewidth]{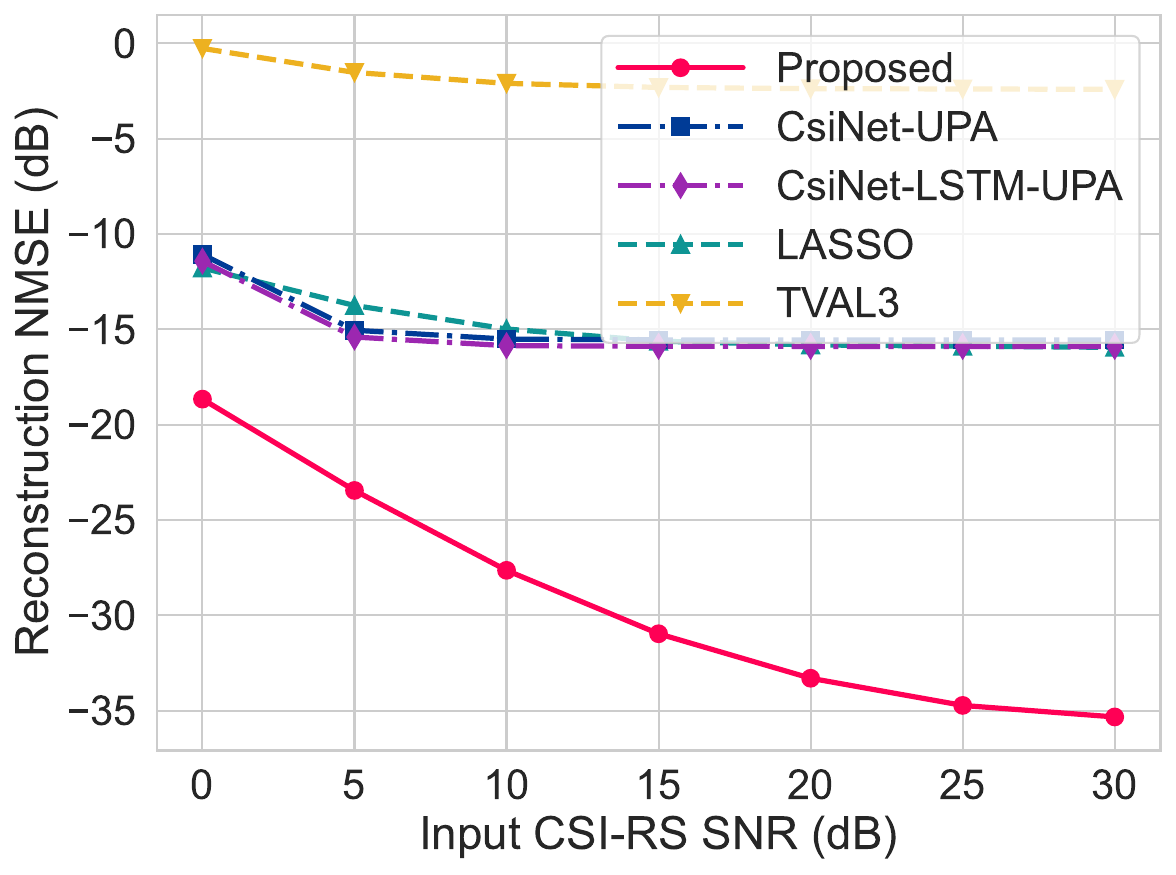}\label{subfig:noise_sionna}}
    \caption{Impact of CSI-RS estimation noise at the receiver. The reconstruction NMSE is evaluated against varying input CSI-RS SNRs for (a) UMa, (b) InF, and (c) CDL-A datasets.}
    \label{fig:csirs_noise_impact}
\end{figure*}

In practical deployments, the downlink channel must be estimated at the UE using pilot signals such as CSI-RS, which introduces estimation noise.
To evaluate the robustness of the proposed framework, we utilize the noisy observed CFR matrix $\mathbf{H}_\text{obs}$ as the input to the compressor.
As shown in Fig.~\ref{fig:csirs_noise_impact}, the TAP architecture acts as an effective denoiser, preserving high reconstruction quality even at low input CSI-RS SNRs.
Notably, the two Sionna RT results (indoor and outdoor) exhibit significantly less performance degradation compared to the CDL-A results.
This enhanced robustness is attributed to two fundamental mechanisms: array averaging and temporal history.

First, by processing the spatial array elements as input channels and collapsing the spatial dimensions to estimate shared delays, the network performs a massive spatial averaging operation.
Independent thermal noise across the antennas cancels out, while the true physical delays add up constructively, drastically boosting the effective SNR.
Second, the network aggregates a temporal context window of past CFRs, acting as a guided denoiser.
Because estimation noise is highly random and uncorrelated over time, whereas physical scattering paths exhibit continuous, predictable trajectories, the 1D-CNN naturally separates temporally correlated signals from independent white noise.
To isolate and confirm the impact of this temporal history, the CDL-A evaluation explicitly uses a temporal window length of 1.
Serving as an ablation study, the steeper degradation of the CDL-A results in Fig.~\ref{fig:csirs_noise_impact}(c) demonstrates that while array averaging provides a robust baseline, the inclusion of temporal history data is a primary driver for the superior noise resilience observed in the Sionna RT environments.

\section{Discussion: Near-Field XL-MIMO Limitations}

To mathematically demonstrate why the current TAP framework is strictly bounded by the far-field assumption, we examine the phase relationship across the wideband spectrum.
In a near-field XL-MIMO regime, the spherical wavefront causes the absolute time of arrival to vary across the spatial array.
For a given multipath component $l$, the true wideband channel at antenna element $m$ and subcarrier $k$ can be modeled as:
\begin{equation}
    [\mathbf{h}_{l,m}]_k = \alpha_{l,m} \exp\left(-j 2\pi f_k (\tau_l + \Delta\tau_{l,m})\right) \label{eq:true_near_field}
\end{equation}
where $\tau_l$ is the reference arrival delay at the array center, and $\Delta\tau_{l,m}$ represents the antenna-dependent delay variation caused by the spherical wavefront curvature.

The TAP framework enforces joint-spatial sparsity by utilizing a DelayNet to extract a single, shared delay estimate $\hat{\tau}_l$ for the entire array.
Consequently, the constructed dictionary basis vector $\mathbf{v}(\hat{\tau}_l) \in \mathbb{C}^{N_{\text{sc}}}$ is spatially invariant in the delay domain:
\begin{equation}
    [\mathbf{v}(\hat{\tau}_l)]_k = e^{-j 2\pi f_k \hat{\tau}_l} \label{eq:st_omp_basis}
\end{equation}

During the spatial reconstruction phase, the LS solver computes an independent complex amplitude for each antenna, denoted as $\hat{x}_{l,m} = |\hat{x}_{l,m}| e^{j\theta_m}$.
This yields the reconstructed channel:
\begin{equation}
    [\mathbf{\hat{h}}_{l,m}]_k = |\hat{x}_{l,m}| \exp(j\theta_m) \exp(-j 2\pi f_k \hat{\tau}_l) \label{eq:ls_reconstruction}
\end{equation}

By isolating the phase terms from \eqref{eq:true_near_field} and \eqref{eq:ls_reconstruction}, and assuming perfect reference delay extraction ($\hat{\tau}_l = \tau_l$), the residual phase error $\epsilon_{m,k}$ at subcarrier $k$ evaluates to:
\begin{equation}
    \epsilon_{m,k} = \underbrace{-2\pi f_k \Delta\tau_{l,m}}_{\text{Near-Field Phase Ramp}} - \underbrace{\theta_m}_{\text{LS Constant Phase}} \label{eq:phase_error}
\end{equation}

Equation \eqref{eq:phase_error} mathematically exposes the structural limitation of 1D spatial-delay dictionaries.
Because the true near-field delay variation $\Delta\tau_{l,m}$ is multiplied by the subcarrier frequency $f_k$, it manifests as a frequency-dependent phase slope across the wideband spectrum.
However, the LS solver is mathematically restricted to applying a single, frequency-independent phase rotation $\theta_m$.
A scalar phase shift can only translate the phase response vertically; it cannot alter the slope.
Therefore, while the LS solver can perfectly align the phase at a specific center frequency, the residual error $\epsilon_{m,k}$ will diverge linearly towards the band edges, leading to severe basis mismatch and spatial energy leakage.

\section{Conclusion and Future Work}

This paper proposed TAP: Tap-Assisted Parametric CSI Compression, a one-shot neural framework for massive MIMO FDD CSI feedback.
By replacing the iterative atom selection of classical OMP with a lightweight CNN-based delay predictor and resolving spatial amplitudes via a closed-form least squares solver, TAP bridges the gap between deep learning and CS.
Evaluations across five 3GPP propagation environments demonstrated that TAP achieves a 10.0 to 11.1~dB improvement in NMSE at the UMa environment and 3.13 to 12.22~dB improvement at the InF environment over the CsiNet family, while operating with a much smaller model footprint and largely accelerating inference compared to iterative OMP, meeting sub-millisecond deployment constraints.

While TAP achieves high-fidelity compression in standard Massive MIMO deployments, its current mathematical structure is bounded by the far-field assumption.
Extending the framework to near-field MIMO scenarios is necessary for next-generation communications.
Furthermore, extending the TAP framework to millimeter-wave (FR2) channels, which impose additional constraints due to hybrid beamforming architectures, is a natural next step.
Validating the zero-shot generalization capabilities on real-world, over-the-air channel measurements will be important for bridging the sim-to-real gap and accelerating the deployment of AI-driven FDD massive MIMO systems.

\bibliographystyle{IEEEtran}
\bibliography{refs}

@STRING{IEEE_J_ITS        = "{IEEE} Trans. Intell. Transport. Syst."}

@STRING{IEEE_J_VT         = "{IEEE} Trans. Veh. Technol."}

@STRING{IEEE_J_SPL        = "{IEEE} Signal Processing Lett."}

@STRING{IEEE_J_CST        = "{IEEE} Trans. Contr. Syst. Technol."}

@STRING{IEEE_J_SP         = "{IEEE} Trans. Signal Processing"}

@STRING{IEEE_J_JSAC       = "{IEEE} J. Sel. Areas Commun."}

@STRING{IEEE_J_COM        = "{IEEE} Trans. Commun."}

@STRING{IEEE_J_WCOM       = "{IEEE} Trans. Wireless Commun."}

@STRING{IEEE_J_IT         = "{IEEE} Trans. Inform. Theory"}

@STRING{IEEE_C_ICC        = "Proc. IEEE Int. Conf. Commun. (ICC)"}

@STRING{IEEE_J_CST        = "{IEEE} Commun. Surveys Tuts."}

@STRING{IEEE_C_WCN        = "Proc. {IEEE} Wireless Commun. Netw. Conf. (WCNC)"}

@STRING{NEURIPS           = "Proc. Advances Neural Inf. Process. Syst. (NeurIPS)"}

@STRING{IEEE_C_ASILOMAR_SSC = "Proc. Asilomar Conf. Signal Syst. Comput."}

@STRING{IEEE_J_WCL        = "{IEEE} Wireless Commun. Lett."}

@STRING{IEEE_J_CCN        = "{IEEE} Trans. Cogn. Commun. Netw."}

@INPROCEEDINGS{Kuo12-WCNC,
  author={Kuo, Ping-Heng and Kung, H. T. and Ting, Pang-An},
  booktitle=IEEE_C_WCN, 
  title={Compressive sensing based channel feedback protocols for spatially-correlated massive antenna arrays}, 
  year={2012},
  volume={},
  number={},
  pages={492-497},
  doi={10.1109/WCNC.2012.6214417}}

@ARTICLE{Huang17-Access,
  author={Huang, Wei and Huang, Yongming and Xu, Wei and Yang, Luxi},
  journal={IEEE Access}, 
  title={Beam-Blocked Channel Estimation for {FDD} Massive {MIMO} With Compressed Feedback}, 
  year={2017},
  volume={5},
  number={},
  pages={11791-11804},
  doi={10.1109/ACCESS.2017.2715984}}

@INPROCEEDINGS{Qi15-ICC,
  author={Qi, Chenhao and Huang, Yongming and Jin, Shi and Wu, Lenan},
  booktitle=IEEE_C_ICC, 
  title={Sparse channel estimation based on compressed sensing for massive {MIMO} systems}, 
  year={2015},
  volume={},
  number={},
  pages={4558-4563},
  doi={10.1109/ICC.2015.7249041}}

@ARTICLE{Shen16-TVT,
  author={Shen, Wenqian and Dai, Linglong and Shi, Yi and Shim, Byonghyo and Wang, Zhaocheng},
  journal=IEEE_J_VT, 
  title={Joint Channel Training and Feedback for {FDD} Massive {MIMO} Systems}, 
  year={2016},
  volume={65},
  number={10},
  pages={8762-8767},
  doi={10.1109/TVT.2015.2508033}}

@ARTICLE{He25-TVT,
  author={He, Xuan and Hou, Hongwei and Fang, Tianhao and Wang, Wenjin and Jin, Shi},
  journal=IEEE_J_VT,
  title={Angle-Delay Domain Hybrid Model-Driven and Data-Driven Downlink {CSI} Acquisition for {FDD} Massive {MIMO} Systems}, 
  year={2025},
  volume={74},
  number={1},
  pages={1788-1793},
  doi={10.1109/TVT.2024.3465846}}

@ARTICLE{Wen18-WCL,
  author={Wen, Chao-Kai and Shih, Wan-Ting and Jin, Shi},
  journal=IEEE_J_WCL, 
  title={Deep Learning for Massive {MIMO} {CSI} Feedback}, 
  year={2018},
  volume={7},
  number={5},
  pages={748-751},
  doi={10.1109/LWC.2018.2818160}}

@INPROCEEDINGS{Pati93-ACSSC,
  author={Pati, Y.C. and Rezaiifar, R. and Krishnaprasad, P.S.},
  booktitle=IEEE_C_ASILOMAR_SSC, 
  title={Orthogonal matching pursuit: recursive function approximation with applications to wavelet decomposition}, 
  year={1993},
  volume={},
  number={},
  pages={40-44 vol.1},
  doi={10.1109/ACSSC.1993.342465}}

@ARTICLE{Tropp07-TIT,
  author={Tropp, Joel A. and Gilbert, Anna C.},
  journal=IEEE_J_IT, 
  title={Signal Recovery From Random Measurements Via Orthogonal Matching Pursuit}, 
  year={2007},
  volume={53},
  number={12},
  pages={4655-4666},
  doi={10.1109/TIT.2007.909108}}

@ARTICLE{Wang19-WCL,
  author={Wang, Tianqi and Wen, Chao-Kai and Jin, Shi and Li, Geoffrey Ye},
  journal=IEEE_J_WCL, 
  title={Deep Learning-Based {CSI} Feedback Approach for Time-Varying Massive {MIMO} Channels}, 
  year={2019},
  volume={8},
  number={2},
  pages={416-419},
  doi={10.1109/LWC.2018.2874264}}

@INPROCEEDINGS{Xu21-WOCC,
  author={Xu, Yang and Yuan, Mingqi and Pun, Man-On},
  booktitle="Proc. Wireless Opt. Commun. Conf. (WOCC)", 
  title={Transformer Empowered CSI Feedback for Massive {MIMO} Systems}, 
  year={2021},
  volume={},
  number={},
  pages={157-161},
  doi={10.1109/WOCC53213.2021.9602863}}

@ARTICLE{Cui22-WCL,
  author={Cui, Yaodong and Guo, Aihuang and Song, Chunlin},
  journal=IEEE_J_WCL, 
  title={TransNet: Full Attention Network for {CSI} Feedback in {FDD} Massive {MIMO} System}, 
  year={2022},
  volume={11},
  number={5},
  pages={903-907},
  doi={10.1109/LWC.2022.3149416}}

@INPROCEEDINGS{Lu20-ICC,
  author={Lu, Zhilin and Wang, Jintao and Song, Jian},
  booktitle=IEEE_C_ICC, 
  title={Multi-resolution {CSI} Feedback with Deep Learning in Massive {MIMO} System}, 
  year={2020},
  volume={},
  number={},
  pages={1-6},
  doi={10.1109/ICC40277.2020.9149229}}

@ARTICLE{Choi17-CST,
  author={Choi, Jun Won and Shim, Byonghyo and Ding, Yacong and Rao, Bhaskar and Kim, Dong In},
  journal=IEEE_J_CST, 
  title={Compressed Sensing for Wireless Communications: Useful Tips and Tricks}, 
  year={2017},
  volume={19},
  number={3},
  pages={1527-1550},
  doi={10.1109/COMST.2017.2664421}}

@ARTICLE{Baraniuk10-TIT,
  author={Baraniuk, Richard G. and Cevher, Volkan and Duarte, Marco F. and Hegde, Chinmay},
  journal=IEEE_J_IT, 
  title={Model-Based Compressive Sensing}, 
  year={2010},
  volume={56},
  number={4},
  pages={1982-2001},
  doi={10.1109/TIT.2010.2040894}}

@ARTICLE{Ji08-TSP,
  author={Ji, Shihao and Xue, Ya and Carin, Lawrence},
  journal=IEEE_J_SP, 
  title={Bayesian Compressive Sensing}, 
  year={2008},
  volume={56},
  number={6},
  pages={2346-2356},
  doi={10.1109/TSP.2007.914345}}

@ARTICLE{Zeng21-TCCN,
  author={Zeng, Jun and Sun, Jinlong and Gui, Guan and Adebisi, Bamidele and Ohtsuki, Tomoaki and Gacanin, Haris and Sari, Hikmet},
  journal=IEEE_J_CCN, 
  title={Downlink {CSI} Feedback Algorithm With Deep Transfer Learning for {FDD} Massive {MIMO} Systems}, 
  year={2021},
  volume={7},
  number={4},
  pages={1253-1265},
  doi={10.1109/TCCN.2021.3084409}}

@INPROCEEDINGS{Xiao23-ICC,
  author={Xiao, Han and Tian, Wenqiang and Liu, Wendong and Zhang, Zhi and Shi, Zhihua and Guo, Li and Shen, Jia},
  booktitle=IEEE_C_ICC, 
  title={A Knowledge-Driven Meta-Learning Method for {CSI} Feedback}, 
  year={2023},
  volume={},
  number={},
  pages={4138-4143},
  doi={10.1109/ICC45041.2023.10279313}}

@inproceedings{Chen26-ICCIP,
author = {Chen, Jiahui and Gu, Xinyu and Li, Haozhen and Liu, Zhenyu},
title = {Unsupervised Adversarial Domain Adaptation for {CSI} Feedback in Codeword Feature Space},
year = {2026},
isbn = {9798400721922},
doi = {10.1145/3784833.3784918},
booktitle = "Proc. Int. Conf. Commun. Inf. Process.",
pages = {269–273},
numpages = {5},
}

@Article{Zhong20-Sensors,
AUTHOR = {Zhong, Shida and Feng, Haogang and Zhang, Peichang and Xu, Jiajun and Luo, Huancong and Zhang, Jihong and Yuan, Tao and Huang, Lei},
TITLE = {Deep Learning Based Antenna Selection for {MIMO} {SDR} System},
JOURNAL = {Sensors},
VOLUME = {20},
YEAR = {2020},
NUMBER = {23},
ARTICLE-NUMBER = {6987},
PubMedID = {33297398},
ISSN = {1424-8220},
DOI = {10.3390/s20236987}
}

@INPROCEEDINGS{He22-GCWkshps,
  author={He, Ke and Vu, Thang X. and Chatzinotas, Symeon and Ottersten, Björn},
  booktitle="Proc. IEEE Globecom Workshops (GC Wkshps)", 
  title={Learning-Based Joint Channel Prediction and Antenna Selection for Massive {MIMO} with Partial {CSI}}, 
  year={2022},
  volume={},
  number={},
  pages={178-183},
  doi={10.1109/GCWkshps56602.2022.10008768}}

@techreport{3gpp.38.214,
 author = {3GPP},
 day = {23},
 institution = "{3rd Generation Partnership Project (3GPP)}",
 month = {Jun.},
 note = {{V19.4.0}},
 number = {38.214},
 title = "{NR; Physical layer procedures for data}",
 type = {Technical Specification (TS)},
 year = {2026}
}

@ARTICLE{Joo26-TCOM,
  author={Joo, Hosung and Choi, Seungmin and Ryu, Sehyun and Yang, Hyun Jong},
  journal=IEEE_J_COM, 
  title={Compressed-{CSI} Feedback with Near Real-time Domain Adaptation}, 
  year={2026},
  volume={},
  number={},
  pages={1-1},
  doi={10.1109/TCOMM.2026.3698862}}

@ARTICLE{Ju24-TWC,
  author={Ju, Hyungyu and Jeong, Seokhyun and Kim, Seungnyun and Lee, Byungju and Shim, Byonghyo},
  journal=IEEE_J_WCOM, 
  title={Transformer-Assisted Parametric {CSI} Feedback for mmWave Massive {MIMO} Systems}, 
  year={2024},
  volume={23},
  number={12},
  pages={18774-18787},
  doi={10.1109/TWC.2024.3476474}}

@techreport{3gpp.38.901,
 author = {3GPP},
 day = {23},
 institution = "{3rd Generation Partnership Project (3GPP)}",
 month = {Jun.},
 note = {{V19.4.0}},
 number = {38.901},
 title = "{Study on channel model for frequencies from 0.5 to 100 GHz}",
 type = {Technical Report (TR)},
 year = {2026}
}

@misc{hoydis23-arXiv_SionnaRT,
      title={Sionna {RT}: Differentiable Ray Tracing for Radio Propagation Modeling}, 
      author={Jakob Hoydis and Fayçal Aït Aoudia and Sebastian Cammerer and Merlin Nimier-David and Nikolaus Binder and Guillermo Marcus and Alexander Keller},
      year={2023},
      note={\textit{arXiv:2303.11103}},
      archivePrefix={arXiv},
      primaryClass={cs.IT},
}

@misc{hoydis23-arXiv_Sionna,
      title={Sionna: An Open-Source Library for Next-Generation Physical Layer Research}, 
      author={Jakob Hoydis and Sebastian Cammerer and Fayçal Ait Aoudia and Avinash Vem and Nikolaus Binder and Guillermo Marcus and Alexander Keller},
      year={2023},
      note={\textit{arXiv:2203.11854}},
      archivePrefix={arXiv},
      primaryClass={cs.IT},
}

@inproceedings{Izacard19-NeurIPS,
author = {Izacard, Gautier and Mohan, Sreyas and Fernandez-Granda, Carlos},
title = {Data-driven estimation of sinusoid frequencies},
year = {2019},
booktitle = NEURIPS,
articleno = {461},
numpages = {11}
}

@INPROCEEDINGS{Feng23-ICC,
  author={Feng, Yijia and Ye, Chenhui and Li, Ruoyi and Pan, Heng and Korpi, Dani},
  booktitle=IEEE_C_ICC, 
  title={{DDA-Net}: A Discrepancy-Based Domain Adaptation Network for {CSI} Feedback Transferability}, 
  year={2023},
  volume={},
  number={},
  pages={4157-4162},
  doi={10.1109/ICC45041.2023.10278810}}

@ARTICLE{Liu24-TWC,
  author={Liu, Zhenyu and Wang, Li and Xu, Lianming and Ding, Zhi},
  journal=IEEE_J_WCOM, 
  title={Deep Learning for Efficient {CSI} Feedback in Massive {MIMO}: Adapting to New Environments and Small Datasets}, 
  year={2024},
  volume={23},
  number={9},
  pages={12297-12312},
  doi={10.1109/TWC.2024.3390583}}

@ARTICLE{Gao15-TSP,
  author={Gao, Zhen and Dai, Linglong and Wang, Zhaocheng and Chen, Sheng},
  journal=IEEE_J_SP, 
  title={Spatially Common Sparsity Based Adaptive Channel Estimation and Feedback for {FDD} Massive {MIMO}}, 
  year={2015},
  volume={63},
  number={23},
  pages={6169-6183},
  doi={10.1109/TSP.2015.2463260}}

@ARTICLE{Liang20-TVT,
  author={Liang, Peizhe and Fan, Jiancun and Shen, Wenhan and Qin, Zhijin and Li, Geoffrey Ye},
  journal=IEEE_J_VT, 
  title={Deep Learning and Compressive Sensing-Based {CSI} Feedback in {FDD} Massive {MIMO} Systems}, 
  year={2020},
  volume={69},
  number={8},
  pages={9217-9222},
  doi={10.1109/TVT.2020.3004842}}

@INPROCEEDINGS{Sattari25-SPAWC,
  author={Sattari, Mehdi and Gündüz, Deniz and Svensson, Tommy},
  booktitle="Proc. IEEE. Int. Workshop Signal Process. Artif. Intell. Wireless Commun. (SPAWC)", 
  title={Dynamically Fine-Tuned Neural Compressor for {FDD} Massive {MIMO} {CSI} Feedback}, 
  year={2025},
  volume={},
  number={},
  pages={1-5},
  doi={10.1109/SPAWC66079.2025.11143260}}

@misc{Zhang26-arXiv,
      title={Towards {CSI}-Native Foundation Models: A Channel-Adaptive Roadmap for {6G}}, 
      author={Chenyu Zhang and Xinchen Lyu and Chenshan Ren and Shuhan Liu and Qimei Cui},
      year={2026},
      note={\textit{arXiv:2606.20670}},
      archivePrefix={arXiv},
      primaryClass={cs.LG},
}

@ARTICLE{Liao25-TITS,
  author={Liao, Yong and Luo, Yu},
  journal=IEEE_J_ITS, 
  title={{DD}-TransNet-Based {CSI} Feedback in Imperfect Channel Estimation in {MIMO-OTFS} Systems for {V2I} Scenarios}, 
  year={2025},
  volume={26},
  number={11},
  pages={18839-18853},
  doi={10.1109/TITS.2025.3590262}}

@ARTICLE{Kang22-JSAC,
  author={Kang, Kai and Hu, Qiyu and Cai, Yunlong and Yu, Guanding and Hoydis, Jakob and Eldar, Yonina C.},
  journal=IEEE_J_JSAC, 
  title={Mixed-Timescale Deep-Unfolding for Joint Channel Estimation and Hybrid Beamforming}, 
  year={2022},
  volume={40},
  number={9},
  pages={2510-2528},
  doi={10.1109/JSAC.2022.3191124}}

@misc{Alkhateeb19-arXiv,
      title={{DeepMIMO}: A Generic Deep Learning Dataset for Millimeter Wave and Massive {MIMO} Applications}, 
      author={Ahmed Alkhateeb},
      year={2019},
      note={\textit{arXiv:1902.06435}},
      archivePrefix={arXiv},
      primaryClass={cs.IT},
}

@ARTICLE{Xi20-TSP,
  author={Xi, Feng and Xiang, Yijian and Chen, Shengyao and Nehorai, Arye},
  journal=IEEE_J_SP, 
  title={Gridless Parameter Estimation for One-Bit {MIMO} Radar With Time-Varying Thresholds}, 
  year={2020},
  volume={68},
  number={},
  pages={1048-1063},
  doi={10.1109/TSP.2020.2970343}}

@ARTICLE{Pali22-SPL,
  author={Pali, Marie-Christine and Ruetz, Simon and Schnass, Karin},
  journal=IEEE_J_SPL, 
  title={Average Performance of {OMP} and Thresholding Under Dictionary Mismatch}, 
  year={2022},
  volume={29},
  number={},
  pages={1077-1081},
  doi={10.1109/LSP.2022.3167313}}

@INPROCEEDINGS{Du24-ICSPCC,
  author={Du, Jiaqi and Yan, Yongsheng and Li, Xiangxiang},
  booktitle="Proc. IEEE Int. Conf. Signal Process. Commun. Comput (ICSPCC)", 
  title={A Low-Complexity Orthogonal Matching Pursuit Algorithm Based on Multi-Scale Multi-Lag Underwater Acoustic Channels}, 
  year={2024},
  volume={},
  number={},
  pages={1-6},
  doi={10.1109/ICSPCC62635.2024.10770462}}

@INPROCEEDINGS{Oyerinde23-ICSPCS,
  author={Oyerinde, Olutayo Oyeyemi and Flizikowski, Adam and Marciniak, Tomasz},
  booktitle="Proc. Int. Conf. Signal Process. Commun. Syst. (ICSPCS)", 
  title={Remodelled and Reduced complexity-{OMP}-based Channel Estimation Schemes for Intelligent Reflecting Surface-Aided Millimeter Wave Systems}, 
  year={2023},
  volume={},
  number={},
  pages={1-5},
  doi={10.1109/ICSPCS58109.2023.10261140}}

@INPROCEEDINGS{Yang19-MLSP,
  author={Yang, Qianqian and Mashhadi, Mahdi Boloursaz and Gündüz, Deniz},
  booktitle="Proc. IEEE Int. Workshop Mach. Learn. Signal Process (MLSP)", 
  title={Deep Convolutional Compression For Massive {MIMO} {CSI} Feedback}, 
  year={2019},
  volume={},
  number={},
  pages={1-6},
  doi={10.1109/MLSP.2019.8918798}}

@Article{Li13-TVAL3,
journal="Comput. Optim. Appl.",
author={Chengbo Li and Wotao Yin and Hong Jiang and Yin Zhang},
title={An efficient augmented Lagrangian method with applications to total variation minimization},
year={2013},
month={December},
pages={507-530},
volume={56},
number={3},
doi={10.1007/s10589-013-9576-1},
}

@techreport{3gpp.38.101.1,
 author = {3GPP},
 day = {7},
 institution = "{3rd Generation Partnership Project (3GPP)}",
 month = {Jul.},
 note = {{V20.0.0}},
 number = {38.101-1},
 title = "{NR; User Equipment (UE) radio transmission and reception; Part 1: Range 1 Standalone}",
 type = {Technical Specification (TS)},
 year = {2026}
}

@ARTICLE{Wang26-TWC,
  author={Wang, Haoyu and Sun, Zhi and Han, Shuangfeng and Wang, Xiaoyun and Wang, Zhaocheng},
  journal=IEEE_J_WCOM, 
  title={Generalizable Learning for Massive {MIMO} {CSI} Feedback in Unseen Environments}, 
  year={2026},
  volume={25},
  number={},
  pages={12514-12530},
  doi={10.1109/TWC.2026.3664171}}

\vfill

\end{document}